\pdfoutput=1 
\documentclass[preprint2]{aastex6}
\usepackage[T1]{fontenc}
\usepackage{lmodern}
\usepackage{graphicx}
\usepackage{amsmath}
\usepackage{amsfonts}
\usepackage{amssymb}

\newcommand{\figfactor}{1.0}

\newcommand{\figfactortwo}{1.0}
\newcommand{\figfactorthree}{1.0}
\newcommand{\figfactorfour}{1.0}

\newcommand{\Tcal}{{\mathcal{T}}}

\begin{document}

\title{Observational Signatures of Mass-Loading in Jets Launched by 
       Rotating Black Holes}
\shorttitle{}

\author{Michael O' Riordan\altaffilmark{1}$^\star$} 
\author{Asaf Pe'er\altaffilmark{1}} 
\author{Jonathan C. McKinney\altaffilmark{2}}

\altaffiltext{1}{Physics Department, University College Cork, Cork, Ireland}
\altaffiltext{2}{Department of Physics and Joint Space-Science Institute,
    University of Maryland, College Park, MD 20742, USA}

\email{$^\star$michael\_oriordan@umail.ucc.ie}

%
%
%

\shortauthors{O' Riordan, Pe'er, \& McKinney}

\begin{abstract}

It is widely believed that relativistic jets in X-ray binaries and 
active-galactic nuclei are powered by the rotational energy of black holes.
This idea is supported by general-relativistic magnetohydrodynamic (GRMHD) 
simulations of accreting black holes, which demonstrate efficient energy 
extraction via the Blandford-Znajek mechanism.
However, due to uncertainties in the physics of mass-loading, and the 
failure of GRMHD numerical schemes in the highly-magnetized funnel region, 
the matter content of the jet remains poorly constrained.
We investigate the observational signatures of mass-loading in the funnel by 
performing general-relativistic radiative 
transfer calculations on a range of 3D GRMHD simulations of 
accreting black holes. 
We find significant observational differences between cases in which 
the funnel is empty and cases where the funnel is filled with plasma,
particularly in the optical and X-ray bands.
In the context of Sgr A*, current spectral data constrains the jet 
filling only if the black hole is rapidly rotating with $a\gtrsim0.9$. 
In this case, the limits on the infrared flux disfavour a strong contribution
from material in the funnel.
We comment on the implications of our models for interpreting future Event
Horizon Telescope observations.
We also scale our models to stellar-mass black holes, and discuss their
applicability to the low-luminosity state in X-ray binaries.

\end{abstract}

\section{Introduction}
\label{sec:intro}

Relativistic jets are a ubiquitous phenomenon. They have been observed across a 
range of 
accreting black hole systems spanning more than 8 orders of magnitude in mass 
-- from stellar-mass black holes in X-ray binaries (XRBs), to supermassive 
black holes in active galaxies.
The Blandford-Znajek (BZ) process \citep{BZ77}, in which rotational energy is
extracted electromagnetically from a Kerr black hole, is widely regarded as
a plausible mechanism for driving these jets.
In a force-free black hole magnetosphere, the BZ model 
predicts that energy is extracted from the black hole at a rate 
$P_\text{BZ}=\kappa\,\Phi^2\,\Omega_\text{H}^2\,/4\pi c$
\citep{BZ77,TNM10}. Here, $\kappa$ is a dimensionless number which 
depends on 
the magnetic field geometry, $\Phi$ is the magnetic flux threading the horizon,
$\Omega_H=ac/2r_\text{H}$ is the angular velocity of the horizon, 
$a$ is the dimensionless black hole spin,  
$r_\text{H}=\left(1+\sqrt{1-a^2}\right) r_g$ is the horizon radius,
and $r_g=GM/c^2$ is the gravitational radius.
The expected BZ jet power therefore depends strongly on the black hole spin, as
well as the properties of the near-horizon magnetic field.

Sophisticated, global general-relativistic magnetohydrodynamic (GRMHD) 
simulations have
largely confirmed the basic predictions of the BZ model.
In particular, \citet{TNM11} and \citet{MTB12} demonstrated jet-launching 
with efficiencies exceeding $100\%$, meaning that more energy flows out of the 
black hole than flows in. 
Such high efficiencies are only possible if enough ordered vertical magnetic 
flux can accumulate near the horizon.
In this case, the magnetic pressure becomes comparable to the gas pressure,
disrupting the inner accretion flow and forming a ``magnetically arrested disk''
\citep[MAD;][]{N+03}.
By contrast, non-MAD flows \citep[called SANE by ][]{Narayan+12} typically do
not show very efficient energy extraction, even at high black hole spin, due to 
the turbulent, disordered fields at the horizon \citep{McB09}.

In recent years, MAD and SANE GRMHD models have been used extensively to model 
Sgr A*, the extremely 
low-luminosity accreting supermassive black hole at the centre of our Galaxy
\citep{Moscibrodzka+09,Shcherbakov+12,MF13,Moscibrodzka+14,Chan+15a,Chan+15b,
    Ball+16,Ressler+17,Gold+17}.
These studies have largely been motivated by very-long baseline 
interferometric (VLBI) observations with 
the Event Horizon Telescope \citep[EHT;][]{Doeleman+09}, which will soon resolve
structure in Sgr A* on spatial scales comparable to the Schwarzschild radius.
The EHT will also resolve small-scale polarized structure, which carries 
information about the near-horizon magnetic field.
Therefore, the EHT offers an unprecedented opportunity to test theories of
accretion and jet-launching, and possibly even general relativity itself 
via measurements of the black hole shadow 
\citep[e.g.,][and references therein]{Psaltis+15}.

Despite these important advances, significant theoretical uncertainties remain
which hinder a direct comparison between the dynamical models and observations.
In particular, there is considerable uncertainty in the mass-loading physics
of BZ jets. 
It is well known that GRMHD codes fail inside the highly-magnetized funnel
\citep{Gammie+03}.
This is because numerical errors accumulate when the ratio of the magnetic energy
density to mass energy density becomes large.
In what follows, we will refer to this ratio as the magnetization $\sigma$.
To keep the numerical scheme stable, GRMHD codes
typically inject matter when $\sigma$ becomes larger than some (rather arbitrary) 
value. This effectively enforces a minimum density in the simulation, 
commonly referred to as a density floor.
Although there are physical processes which may operate to mass-load the funnel,
for example pair cascades \citep{BZ77,LR11,BT15} or photon annihilation 
\citep{Moscibrodzka+11},
the injection of floor material is arbitrary and chosen 
simply to avoid numerical issues. Therefore, the funnel mass and internal energy 
densities are not determined by the GRMHD simulations.

Although the injected floor material has little effect on the dynamics, it can 
affect the resulting spectra and so must be considered when comparing
GRMHD models with observations. 
Depending on the choice of initialisation for the floors, the plasma in the funnel 
might 
be tenuous enough such that it has a negligible contribution to the
spectra. In this case, the jet emission is dominated by the funnel wall or
``jet sheath'' as in \citet{MF13} and \citet{Moscibrodzka+14}.
This ``empty funnel'' situation can also be achieved by simply removing 
floor material
from the funnel during the radiative transport calculation. The material to remove
can be chosen in a number of ways, for example as cells in a large bipolar cone 
\citep{SM13}, cells considered artificially hot or dense relative to their
neighbours \citep{Chan+15a}, or cells with a large value of 
$\sigma$ \citep{O'Riordan+16b,O'RIordan+16a}.

Recently, \citet{Gold+17} argued that the prescription used for treating the 
funnel material could be very important when interpreting future 
observations from the EHT.
In particular, they showed that the black hole shadow can be completely 
obscured in the case of significant emission from the funnel, while the absence 
of strong funnel emission can in fact mimic features of the shadow.
Therefore, in order to test general relativity using EHT observations it will be 
crucial to distinguish between features caused by strong-field gravity and 
those arising from the presence or absence of emitting matter in the jet.

In this work, we investigate the observational effects of mass-loading
in the regime where the funnel remains force-free. That is, we restrict
our analysis to the case where the funnel material is highly-magnetized with 
$\sigma\gtrsim 10$.
In the opposite regime where the inertia of the funnel plasma cannot be neglected 
($\sigma\lesssim 1$), \citet{GL13} showed 
that mass and energy loading of the field lines can strongly suppress or even 
switch off energy extraction from the black hole. 
This case would therefore involve significant modifications to the dynamical
GRMHD models. We will study the observational consequences of this regime
in a future work.

The structure of the paper is as follows.
In Section~\ref{sec:models} we briefly describe our GRMHD models, 
radiative transport code, and prescriptions for treating the electrons 
in the jet. 
In Section~\ref{sec:results} we show the spectra from our GRMHD models and 
describe the observational effects of mass-loading the funnel.
In Section~\ref{sec:discussion} we summarise and discuss out findings.
Throughout the paper we use units where $G=c=1$, which implies that the 
gravitational radius $r_g$ and light-crossing time $t_g=r_g/c$ become 
$r_g=t_g=M$. We will occasionally reintroduce factors of $c$ for clarity.

\section{Models}
\label{sec:models}
\subsection{GRMHD Simulations}

\begin{figure}
    \centering
    \includegraphics[width=\figfactorthree\linewidth]{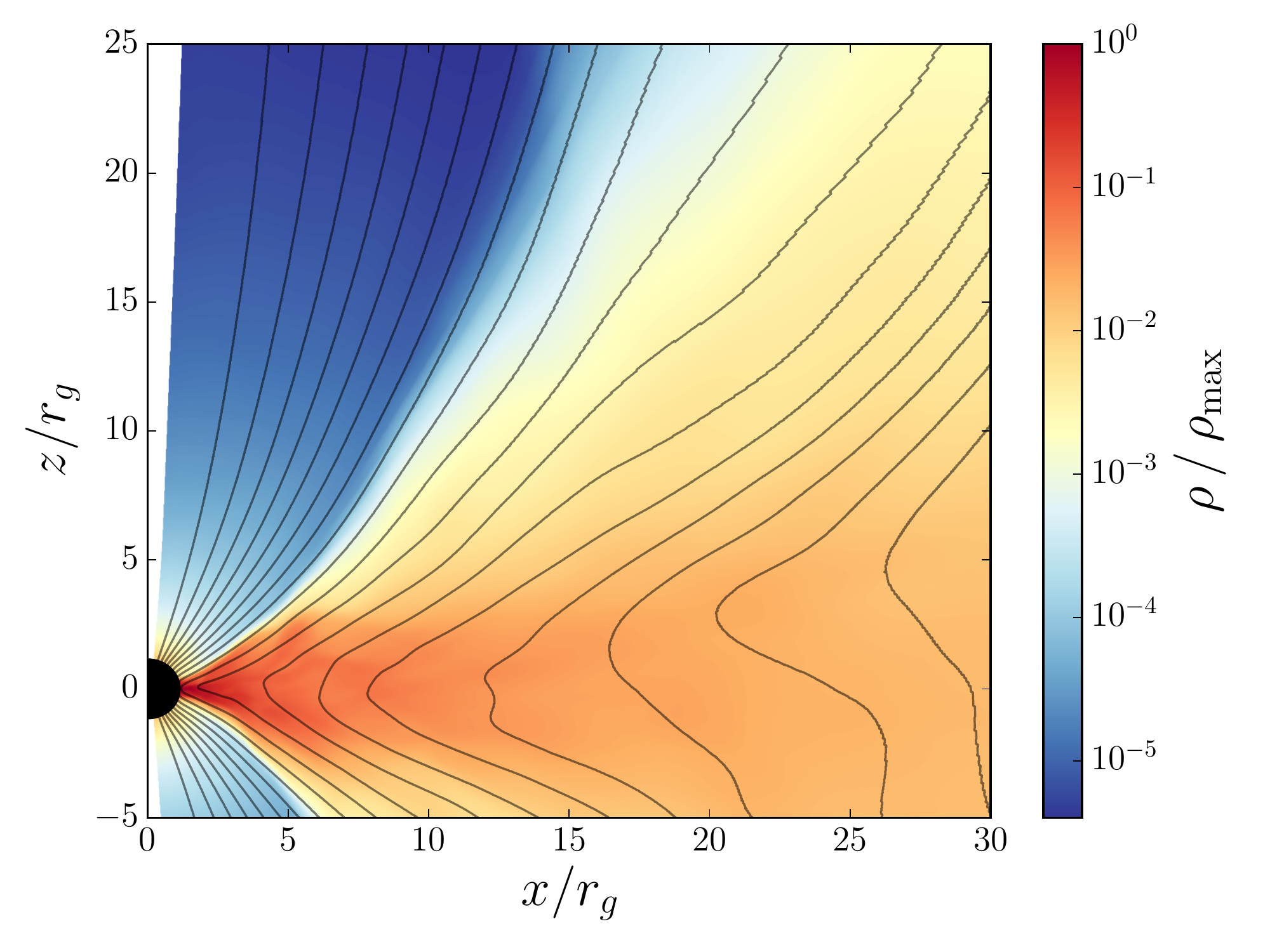}
    \includegraphics[width=\figfactorthree\linewidth]{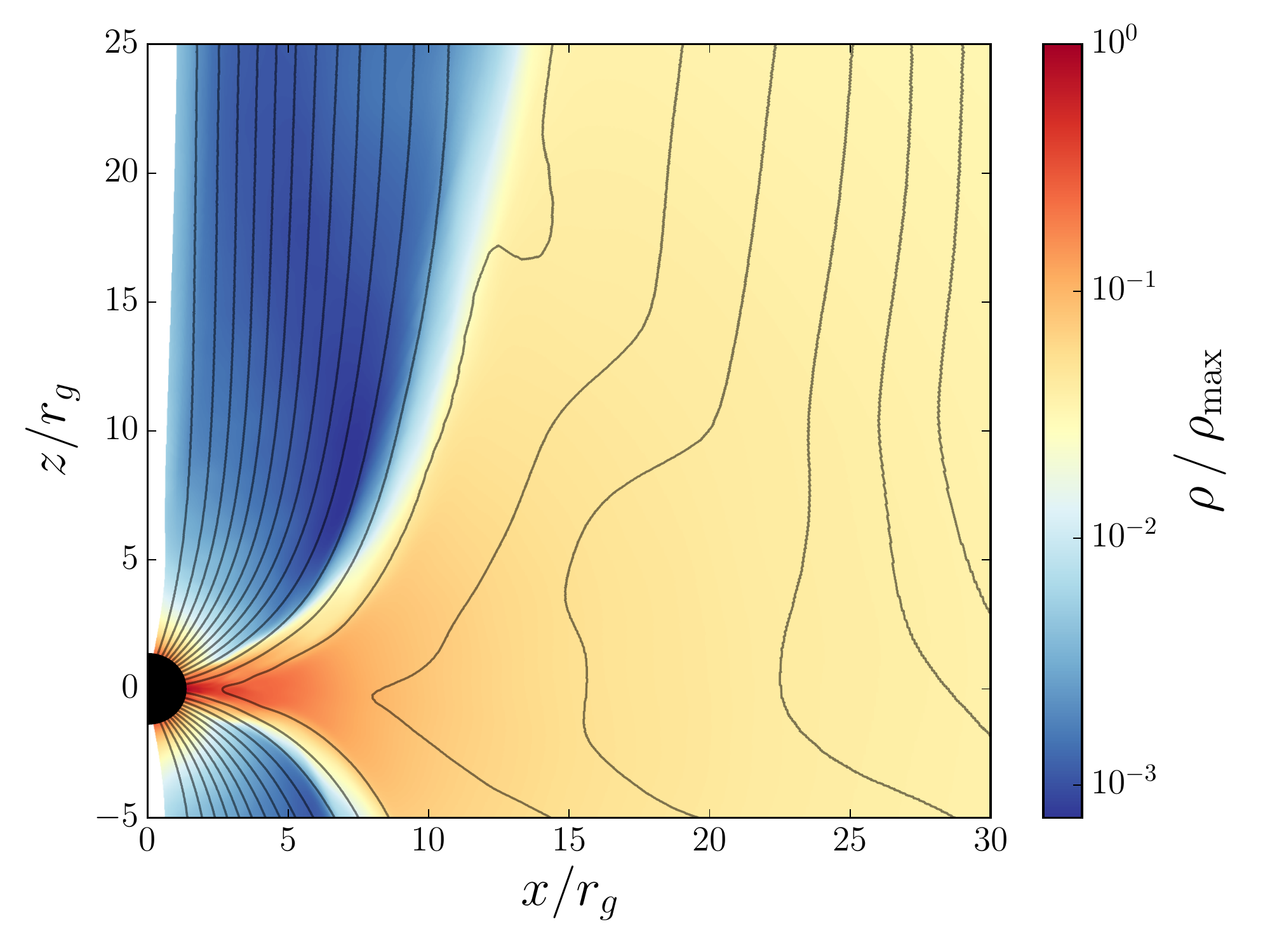}
    \includegraphics[width=\figfactorthree\linewidth]{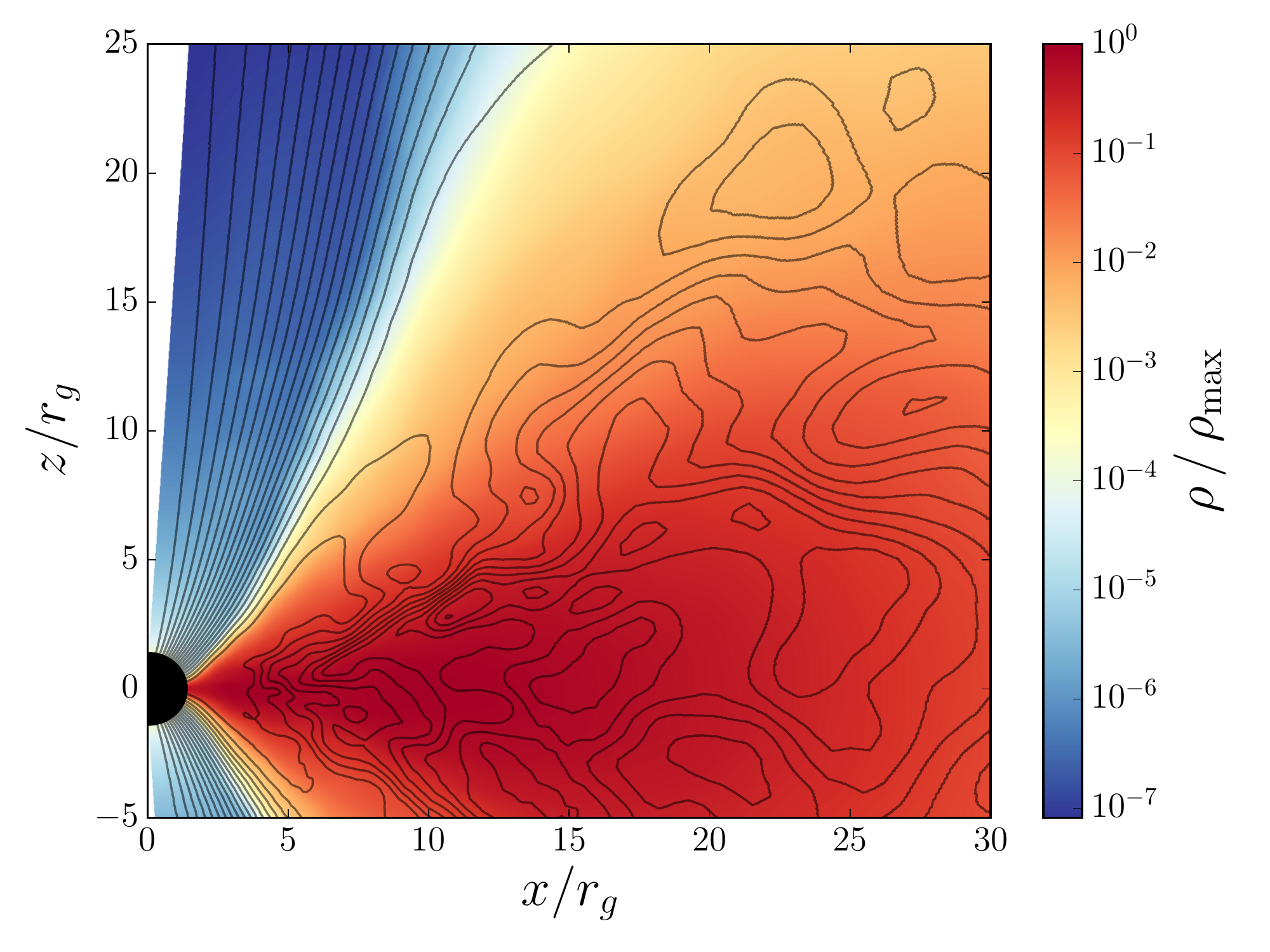}
    \caption{Snapshots of our MAD and SANE GRMHD models. The colour shows the mass density
       and the poloidal magnetic field lines are represented by the black contours.
       The top panel shows the thin-MAD model with $H/R\approx 0.2$ 
       and $a=0.99$. The
       middle panel shows the thick-MAD model with $H/R\approx 1$ and $a=0.9375$. 
       The bottom panel shows the SANE model with $H/R\approx 0.2$ and $a=0.92$.
       Both MAD models have large-scale ordered poloidal fields in the disk and 
       jet.
       The white regions along the $z$-axis correspond to material that has been 
       removed to avoid numerical issues due to coordinate singularities.
       The blue regions roughly correspond to the numerical density floors, 
       which are removed in our ``empty'' funnel models 
       (see the text for details about the floor removal).
}
    \label{fig:harm_models}
\end{figure}

We consider six MAD accretion flows from \citet{TNM11} and \citet{MTB12}, 
and a SANE accretion flow from \citet{McB09}. 
Five of our MAD models have a scale height of $H/R\approx 0.2$ and spins of 
$a=\left\{0.1,\,0.2,\,0.5,\,0.9,\,0.99\right\}$.
These are called A0.1N100, A0.2N100, A0.5N100, A0.9N100, and A0.99N100 
in \citet{MTB12}. We will refer to these as our ``thin-MAD'' models.
We also consider a very geometrically-thick MAD model with $H/R\approx 1$ and a 
spin of $a=0.9375$, called A0.94BfN40 in \citet{MTB12}. We will refer to this
model as ``thick-MAD''. Finally, we consider a SANE model with $H/R\approx 0.2$
and a spin of $a=0.92$, called MB09D in \citet{MTB12}.

In Figure~\ref{fig:harm_models} we show snapshots of our MAD and SANE models.
The colour shows the mass density and the black contours show the structure of 
the poloidal magnetic field (from the $\phi$-integrated vector potential).
The top panel shows the thin-MAD model with $a=0.99$, the middle panel shows the
thick-MAD model with $a=0.9375$, and the bottom panel shows the SANE model with
$a=0.92$. The MAD models have large-scale, ordered poloidal fields in 
the disk and jet, while the disk in the SANE model has a more disordered 
field. 
In all models, we remove material from cells near the poles as coordinate 
singularities
can cause numerical issues here. This is indicated as an excised region along 
the $z$-axis in Figure~\ref{fig:harm_models}.
Detailed descriptions of these models can be found in 
\citet{McB09,TNM11,MTB12,O'Riordan+16b,O'RIordan+16a}.

\subsection{Electron Temperature Prescription}

We calculate the spectra from these models in a post-processing step using a 
general-relativistic radiative 
transport code based on \texttt{grmonty} \citep{Dolence+09}.
We use snapshots from the GRMHD simulations as input, and include 
contributions to the spectra from
synchrotron emission, absorption, and Compton scattering from relativistic 
thermal electrons. 
The mass accretion rates in our low-luminosity target applications of Sgr A* and 
the low/hard state in XRBs are expected to be well below the corresponding 
Eddington rate, which justifies treating the radiation in a post-processing 
step. The Eddington rate is defined as 
$\dot{M}_\text{Edd}c^2\equiv 10\,L_\text{Edd}
\approx 10^{39} \left(\frac{M}{M_\odot}\right) \text{erg s}^{-1}$
\citep{NM08}. 
Since differences in mass density and magnetization can cause different heating 
and cooling rates for the protons and electrons in the disk and jet
\citep[e.g.,][]{Ressler+15,Foucart+16}, we specify the proton-to-electron 
temperature ratio $\Tcal\equiv T_p/T_e$ as a
function of the plasma $\beta\equiv p_\text{gas}/p_\text{mag}$.
Here, $p_\text{gas}$ is the thermal pressure of the fluid, and $p_\text{mag}$ is
the magnetic pressure.
In order to maximize the potential contributions from the highly-magnetized 
funnel material, unless otherwise specified, we choose a critical value of 
$\beta_c=0.2$ and set
$\Tcal=\Tcal_\text{disk}=30$ in regions where 
$\beta>\beta_c$, and
$\Tcal=\Tcal_\text{jet}=3$ in regions where $\beta\leq\beta_c$.
For simplicity, we will refer to regions with $\beta\leq\beta_c$ as the 
``jet'', 
and regions with $\beta>\beta_c$ as the ``disk''. In particular, the ``jet'' 
includes both the funnel wall and central funnel matter.
Although the choice of $\beta_c$ is somewhat arbitrary, we find that using 
$\beta_c=0.2$ gives a reasonable distinction between the 
disk and jet, and our results are largely unaffected by small changes in 
$\beta_c$ up to a factor of a few.
We impose a smooth, exponential transition between the temperature ratios in the 
disk and jet by setting
$\Tcal = \Tcal_\text{jet}\, e^{-\beta/\beta_c} + \Tcal_\text{disk} \left(1-e^{-\beta/\beta_c}\right)$.

\subsection{Empty Funnel Prescription}

To maintain numerical stability, GRMHD codes must inject material into
the low-density, highly-magnetized, funnel region. 
In particular, numerical errors accumulate when the magnetization becomes large
$\sigma=b^2/\rho\gg1$. Here,
$b^2=b^\mu b_\mu$, $b^\mu$ is the magnetic four-field, and $\rho$ is the 
rest-mass density. 
The magnetic four-field can be written in terms of the lab-frame 3-field $B^i$ as
$b^\mu=h^\mu_\nu\,B^\nu/u^t$, where $u^\mu$ is the fluid four-velocity and 
$h^\mu_\nu=\delta^\mu_\nu+u^\mu u_\nu$ is a projection tensor.
In our units, the magnetic pressure is related to the magnetization by 
$p_\text{mag} = \sigma \rho / 2$.
The injected floor material roughly 
corresponds to the blue regions in Figure~\ref{fig:harm_models}.
Although the injected numerical
density floors do not affect the dynamics, they can be artificially hot and so
might affect the resulting spectra. 
In \citet{O'Riordan+16b,O'RIordan+16a}, we considered the case where
material from the central regions of the funnel doesn't contribute 
significantly to the observed spectrum. 
That is, we removed the floor material such that the middle of the 
funnel region was empty.
In this work, we follow the same procedure for removing the floor material 
and will refer to the resulting models as ``empty''.
For removing the floors, we set the density to zero
in regions where $\sigma>\sigma_c\left(r\right)$. We use $\sigma_c=20$ at the 
horizon, and linearly interpolate to $\sigma_c=10$ at $r=10\,r_g$. For larger 
radii, we use a fixed value of $\sigma_c=10$. This ensures that the injected floors
are removed, without removing material close to the black hole which 
naturally becomes highly magnetized. Using this prescription, the centre of the 
funnel region is removed while the disk and funnel wall are not affected.
The dashed lines in Figure~\ref{fig:harm_models_filled}, which we will refer to 
as the ``edge'' of the funnel wall, show the regions that
are removed using this prescription.

\subsection{Filled Funnel Prescription}

We also consider the case where the funnel is mass-loaded and will refer 
to these models as ``filled''. When modelling the filled funnel, we restrict 
our attention to the regime in which 
the mass-loading of the jet doesn't affect the magnetic field in the funnel. 
In covariant form, the energy and momentum exchange between an 
electromagnetic field and charged matter can be written as 
$\nabla_\mu T^{\mu\nu}_\text{EM}=-F^{\mu\nu}j_\nu$, where 
$\nabla_\mu$ is the covariant derivative,
$T^{\mu\nu}_\text{EM}=F^{\mu\alpha} F^\nu{}_\alpha
-\frac{1}{4}\,g^{\mu\nu}F_{\alpha\beta}F^{\alpha\beta}$ is the electromagnetic 
stress-energy tensor, $F^{\mu\nu}$ is the
electromagnetic field tensor, $j^\mu$ is the electric four-current density, and
$g_{\mu\nu}$ is the metric.
In the case where the plasma energy-momentum is many orders of magnitude less 
than that of the electromagnetic field, the energy and momentum exchange can be 
neglected. In this case, the electromagnetic stress-energy tensor is conserved 
by itself $\nabla_\mu T^{\mu\nu}_\text{EM}=0$.
Such a situation is referred to as force-free because of the vanishing of 
the Lorentz four-force density $f^\mu=F^{\mu\nu}j_\nu$.
The approximately force-free solution in the funnel will be preserved as long as 
the injected
matter has $\sigma\gg1$ \citep{MG04}. In this regime we can treat the funnel
mass-loading in a post-processing step.
More significant mass-loading with $\sigma\lesssim1$ would affect the fluid 
dynamics and could even quench the BZ jet \citep{GL13}.
In the case of a strongly mass-loaded funnel, the resulting GRMHD solution 
may deviate significantly from the models described here.

Various processes have been proposed which act to fill 
the funnel with electron-positron ($e^\pm$) or electron-proton ($e$-$p$) pairs,
however
the physical mechanism which operates in nature to mass-load jets remains an 
open problem. 
GRMHD simulations typically show the formation of a surface near the 
black hole which separates the inflowing and outflowing plasmas. 
This ``stagnation'' surface is continuously evacuated, resulting in large 
unscreened electric fields. 
Therefore, the stagnation surface might be the location of 
$e^\pm$ pair formation and subsequent acceleration \citep{LR11,BT15}.
Furthermore, depending on the radiation field produced by the inner regions of 
the accretion flow, the funnel might be filled with $e^\pm$ pairs via 
photon annihilation \citep{Moscibrodzka+11}.
While these mechanisms both result in $e^\pm$ jets, there are also 
magnetohydrodynamic processes which might fill the jet with 
$e$-$p$ pairs.
These include magnetic Rayleigh-Taylor 
instabilities in the funnel wall \citep[][and Appendix~\ref{sec:filling}]{MTB12}, 
and magnetic field polarity 
inversions in the disk \citet{Dexter+14}. Both of these processes inject 
matter from the disk into the centre of the funnel. 
In this work, we do not specify a mass-loading mechanism, but 
instead consider the limiting cases of an empty funnel and a funnel filled 
with constant profiles of mass and internal energy density. 
We set the density and internal energy to be as large as possible, while still 
satisfying the force-free condition.
Therefore, we expect the spectra from mass-loaded force-free jets to fall 
between the extremes considered here.

For our filled models, 
we first remove the floor material using the procedure described above, and
then fill the empty funnel cells at each radius with constant mass and internal 
energy densities,
equal to their corresponding values at the edge of the funnel wall (denoted by
the dashed lines in Figure~\ref{fig:harm_models_filled}). 
We then re-scale the material in the funnel and funnel 
wall to conserve energy. In practice, this re-scaling has little effect on the 
resulting spectra. 
Using this procedure, the properties of the plasma in the funnel are determined 
by the self-consistent material in the funnel wall.
The resulting matter distribution in the funnel is in fact similar to
the original floor material shown in Figure~\ref{fig:harm_models}. However,
we choose to manually fill the funnel to avoid any potential 
issues with artificially hot cells, which would otherwise have to be 
checked and removed as in \citet{Chan+15a}.
We show the mass and internal energy density distributions in our empty
and filled models in Figure~\ref{fig:harm_models_filled}.

\newpage
\section{Results}
\label{sec:results}

\subsection{Predictions for Spectra of Sgr A*}
\label{sec:sgr_spectra}

\begin{figure}
    \centering
    \includegraphics[width=\figfactortwo\linewidth]{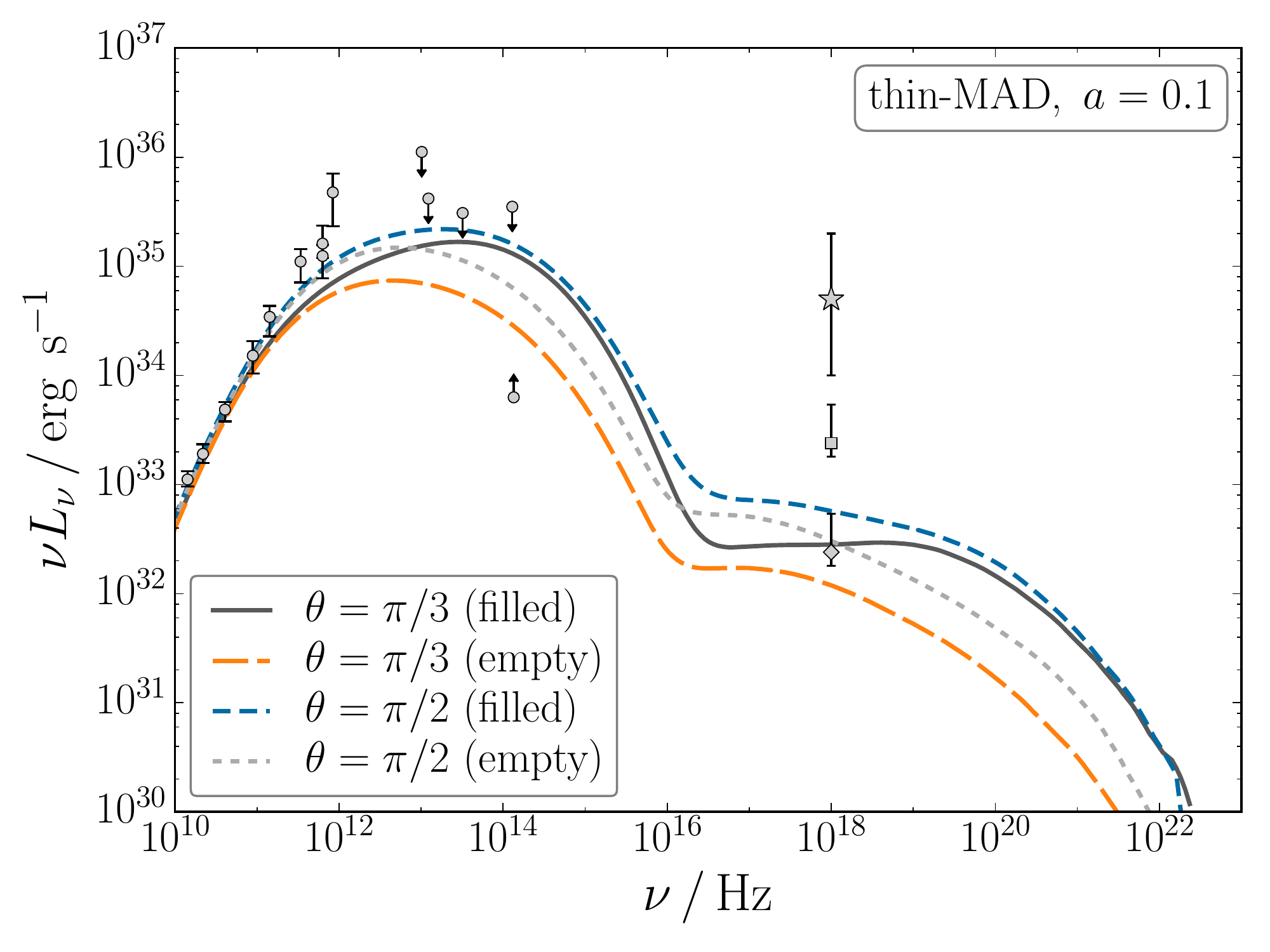}
    \caption{Comparison of the spectra for the empty and filled funnel thin-MAD models with $a=0.1$. 
The radio data points and IR limits are the same as those considered by 
\citet{Chan+15a}. 
The X-ray flux during quiescence is marked by the square data 
point \citep{Baganoff+03}, while the diamond marks $10\%$ of the quiescent 
X-ray flux \citep{Neilsen+13}. 
The range of observed X-ray flares is represented by the 
star \citet{Neilsen+13}.
The radio emission 
originates in the funnel wall and so is not sensitive to the mass-loading of the funnel.
The funnel material primarily contributes to the IR and optical bands, with a corresponding
increase in the synchrotron self-Compton component.
In this low-spin case, both the empty and filled funnel models are largely consistent
with the data.
}
    \label{fig:thinnermad1}
\end{figure}

\begin{figure}
    \centering
    \includegraphics[width=\figfactortwo\linewidth]{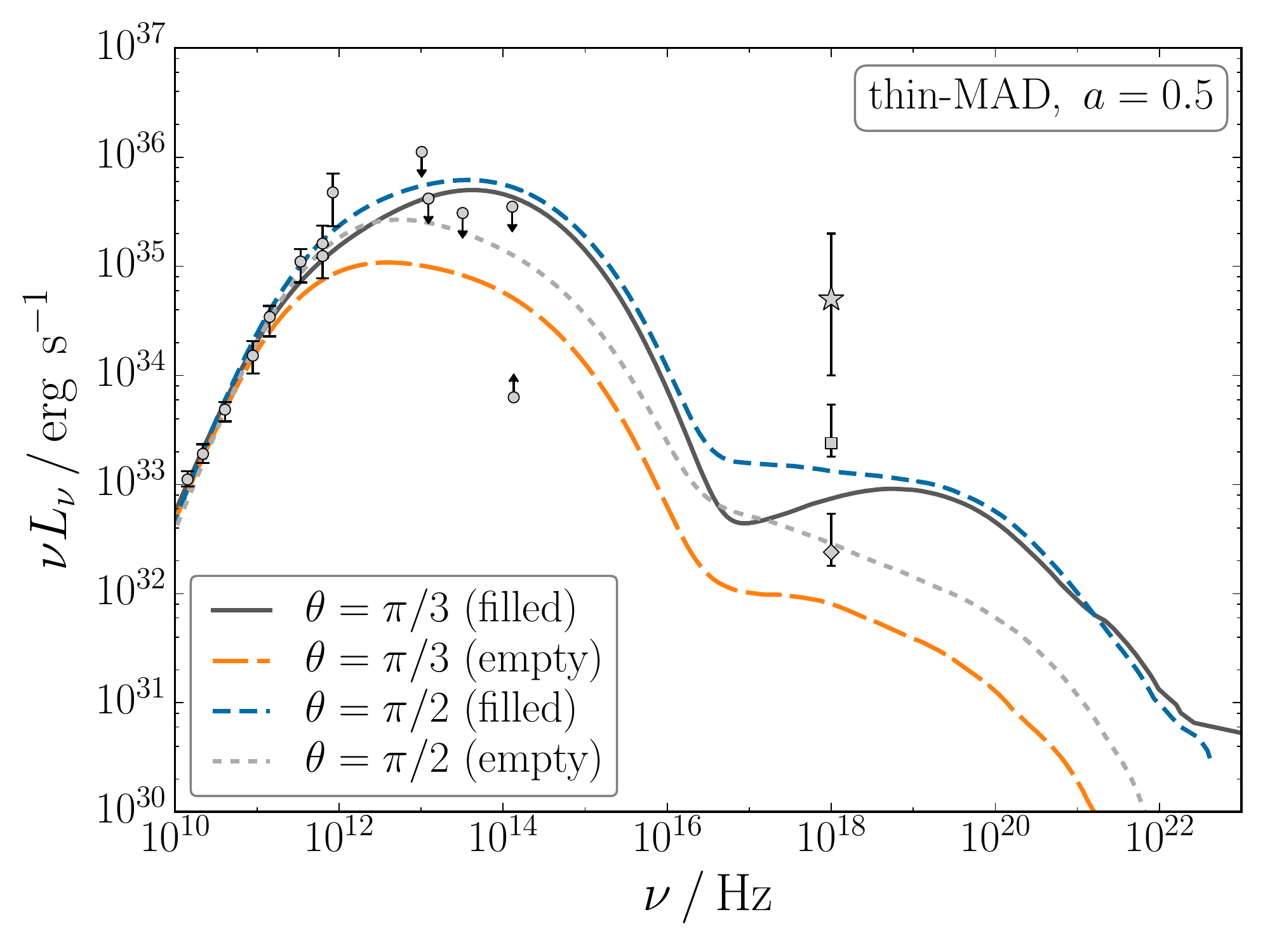}
    \includegraphics[width=\figfactortwo\linewidth]{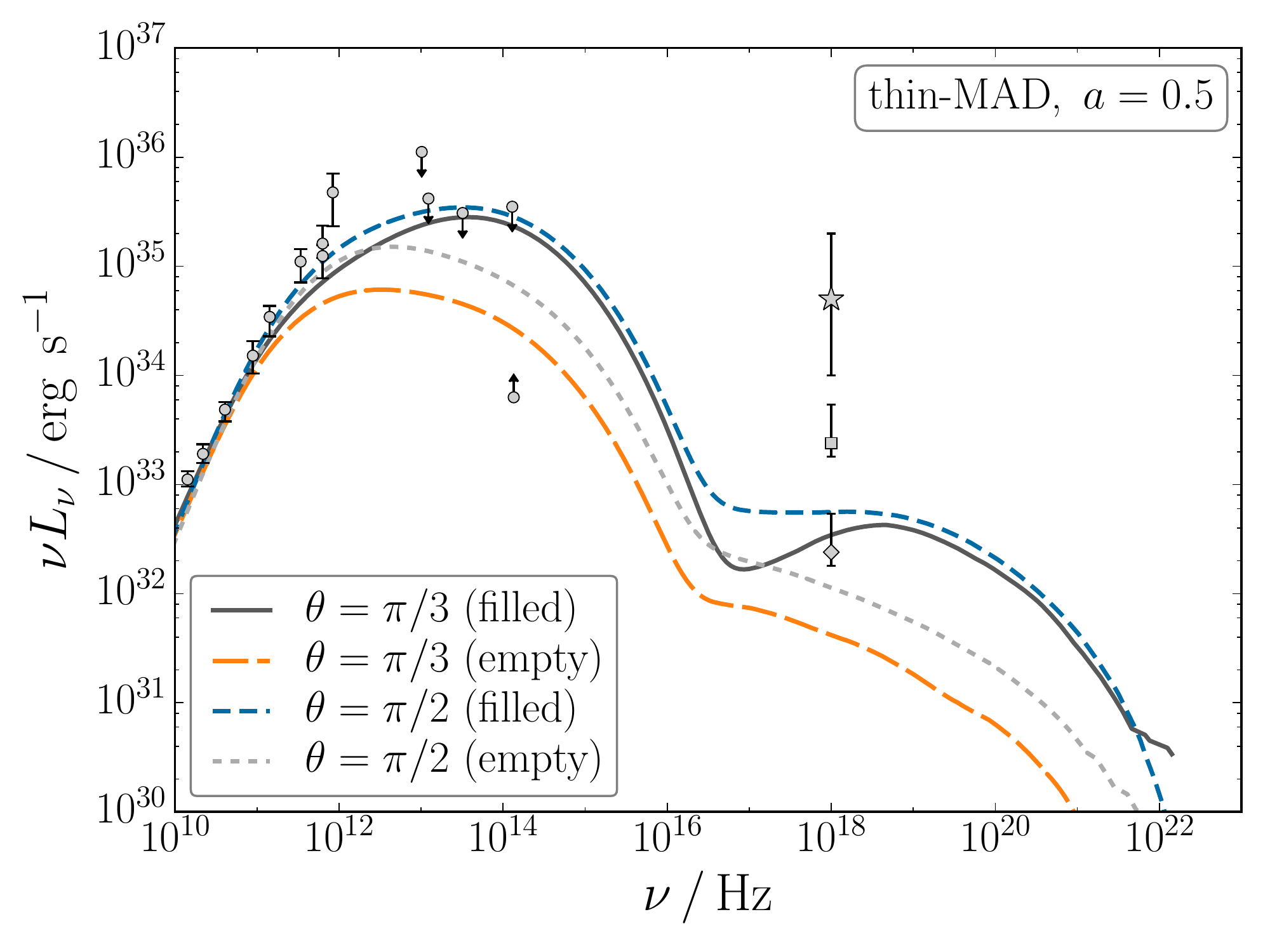}
    \caption{Spectra for the thin-MAD model with $a=0.5$. 
        The spectra are qualitatively
similar to the $a=0.1$ case, but with a larger contribution from the funnel material.
To obtain better fits with the filled model, the accretion rate in the bottom 
panel has been decreased by a factor of $\sim 1.5$ relative to that in the top 
panel.
Although both the empty and filled funnel models are consistent with the data, the 
IR emission in the filled funnel case is close to the maximum flux allowed by observations.
}
    \label{fig:thinnermad5}
\end{figure}

\begin{figure}
    \centering
    \includegraphics[width=\figfactor\linewidth]{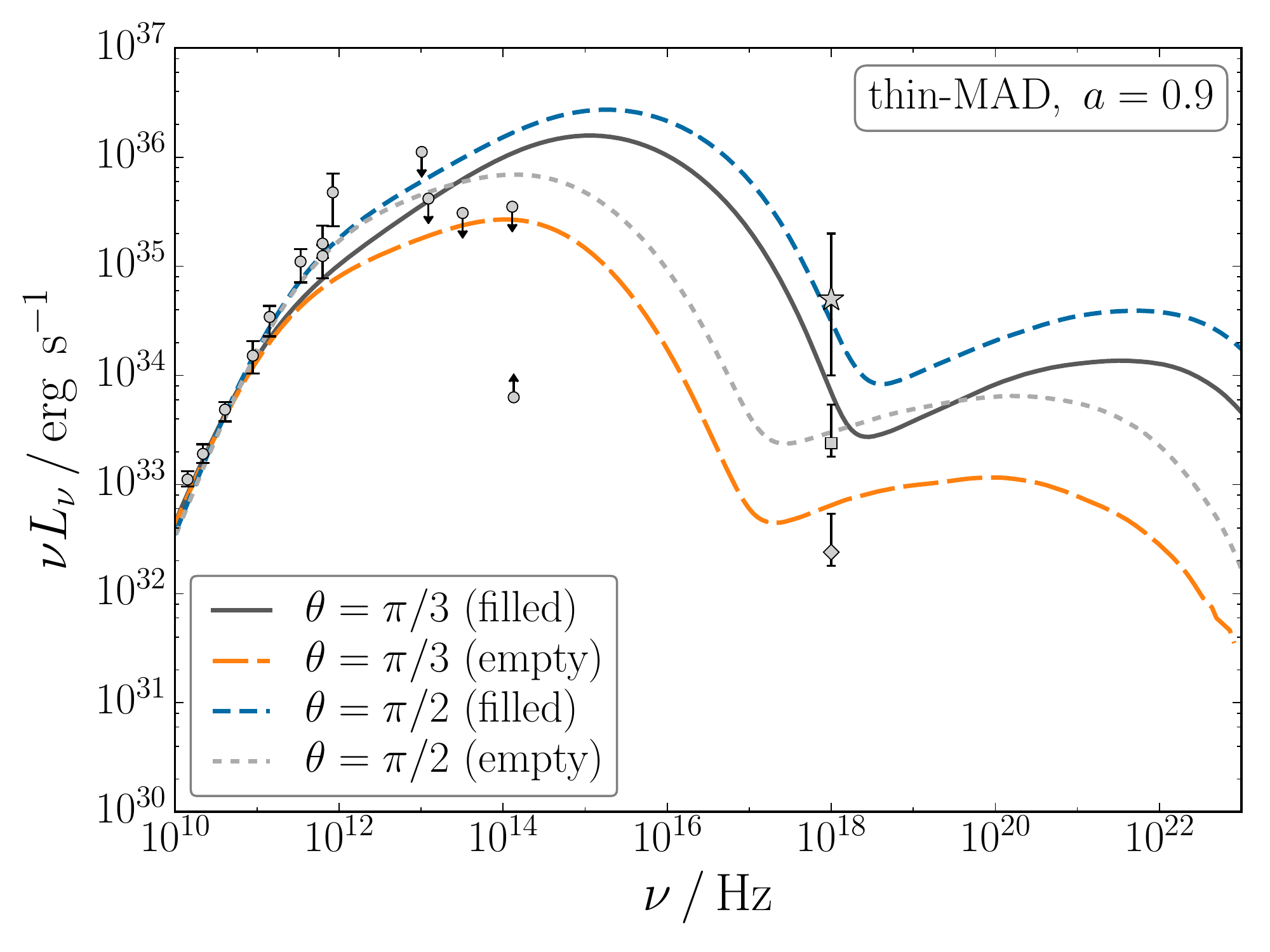}
    \caption{Spectra for the thin-MAD model with $a=0.9$. In this case the IR
limits and X-ray estimates disfavour a filled funnel component. Even the empty 
funnel case is approaching the limits of the observations.
While a lower accretion rate would decrease the IR and X-ray flux towards values
more consistent with the data, the radio flux would then be missed by a large amount.
}
    \label{fig:thinnermad9}
\end{figure}

\begin{figure}
    \centering
    \includegraphics[width=\figfactor\linewidth]{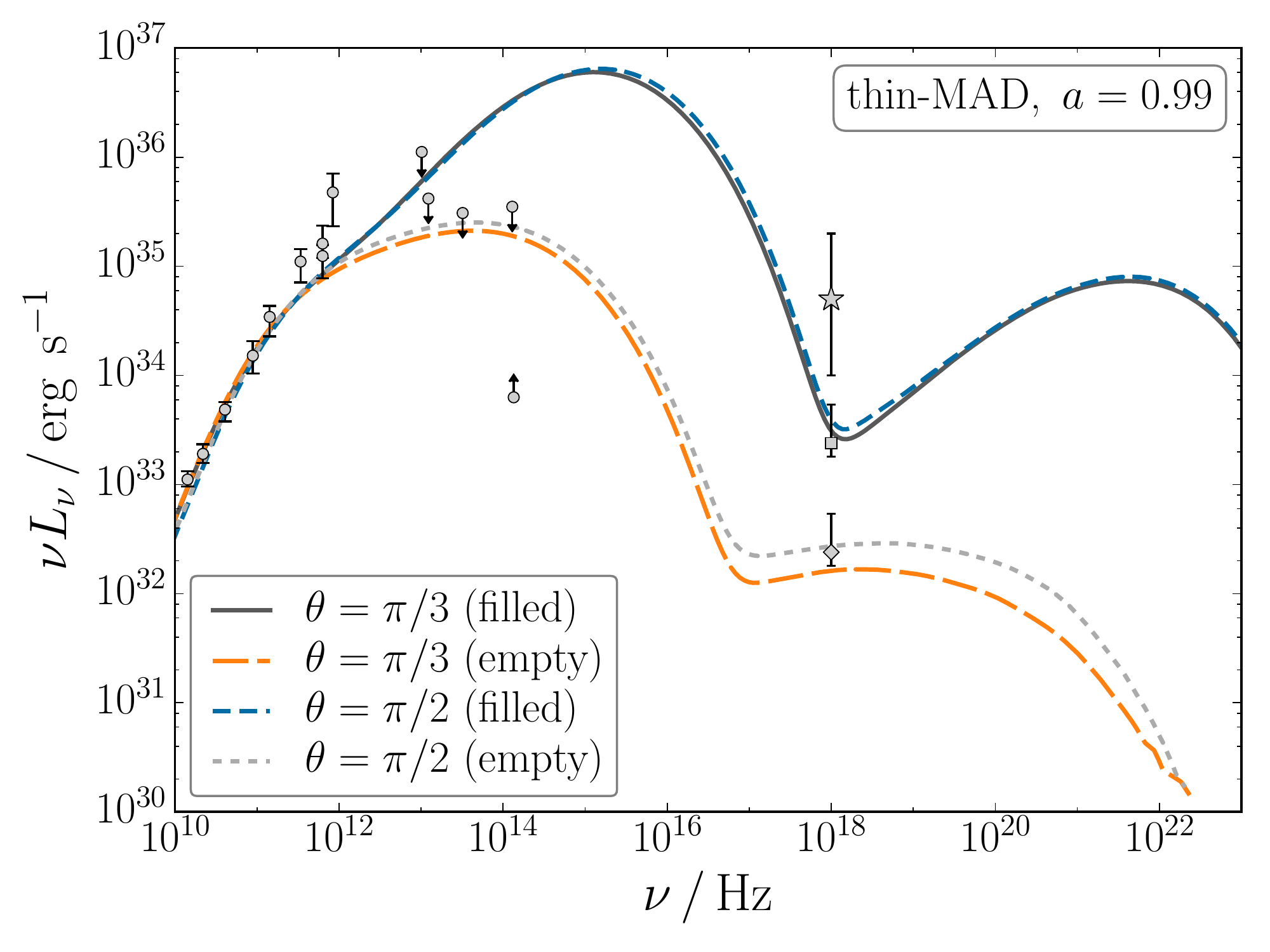}
    \caption{Spectra for the thin-MAD model with $a=0.99$. 
As with the $a=0.9$ case, the IR and X-ray data disfavour the filled funnel
model. Furthermore, fitting the empty funnel model to the data requires suppressing
the emission from close to the horizon by increasing the proton-to-electron 
temperature ratio of the inflowing material.
}
    \label{fig:thinnermad99hc}
\end{figure}

\begin{figure}
    \centering
    \includegraphics[width=\figfactor\linewidth]{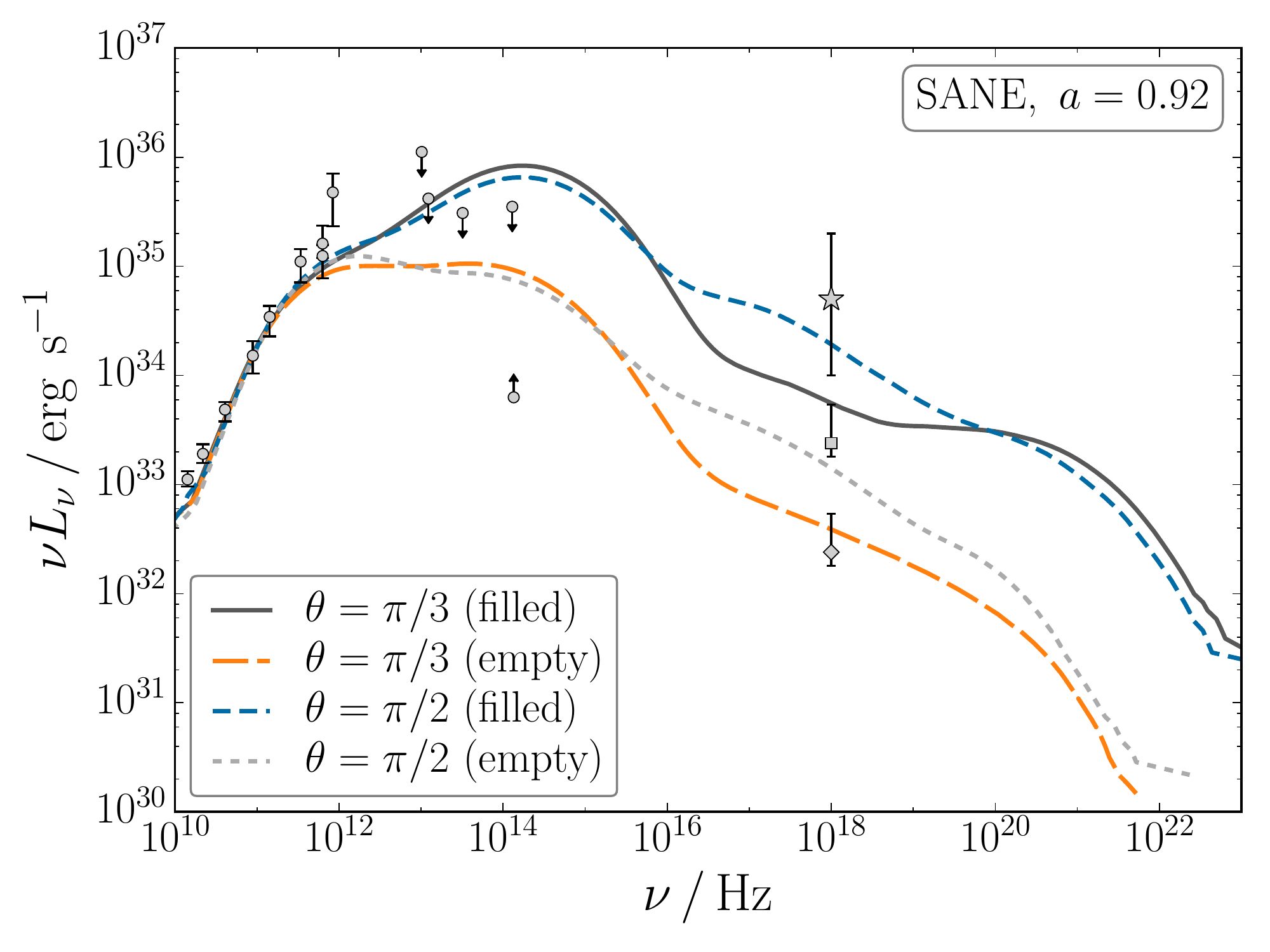}
    \caption{Spectra for the SANE model with $a=0.92$. As in the high-spin MAD models,
the IR limits disfavour models with strong funnel emission. }
    \label{fig:nonmad}
\end{figure}

\begin{figure}
    \centering
    \includegraphics[width=\figfactortwo\linewidth]{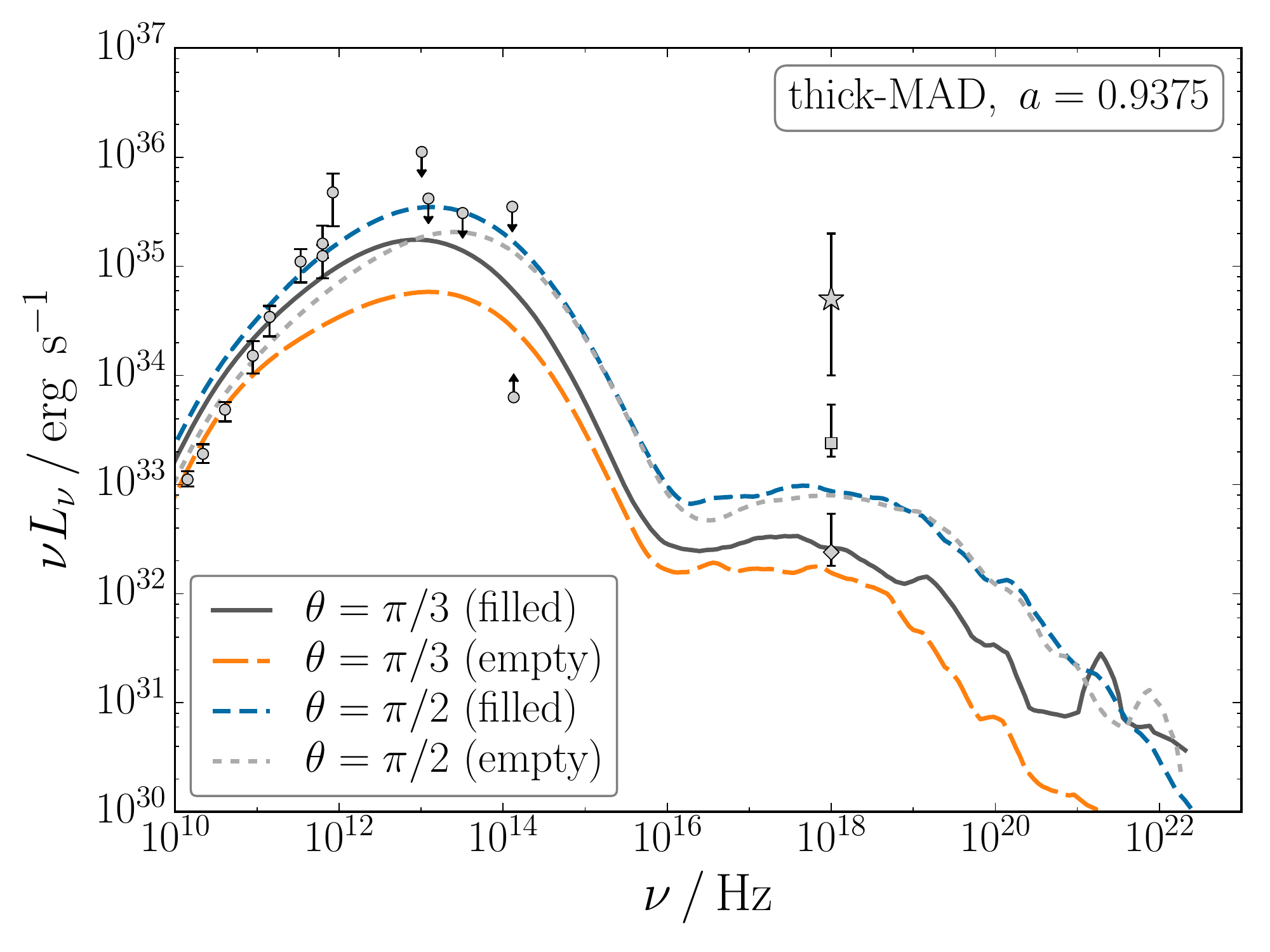}
    \includegraphics[width=\figfactortwo\linewidth]{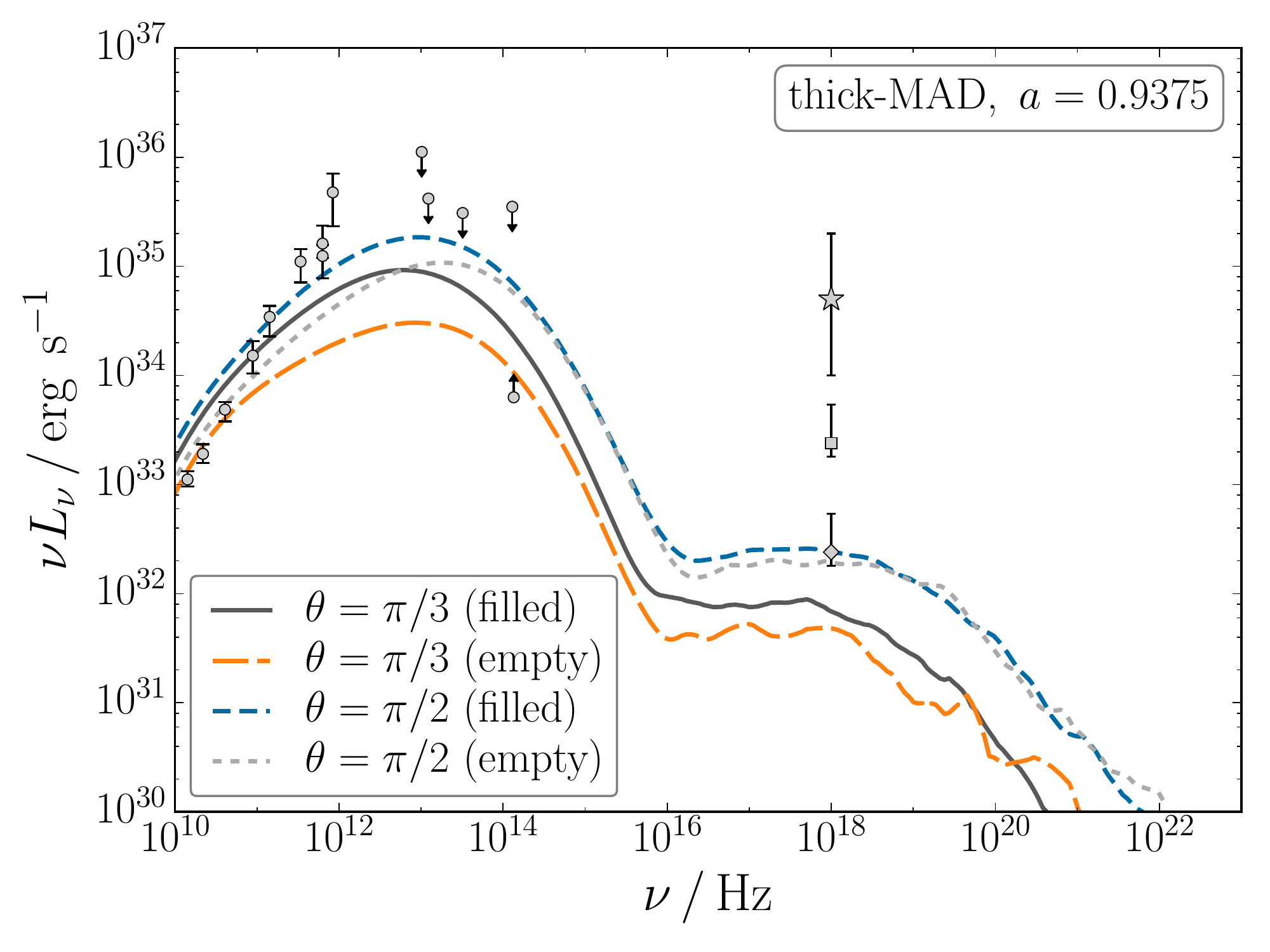}
    \caption{Spectra for the geometrically-thick ($H/R\approx 1$) MAD model with 
$a=0.9375$. The accretion rate in the top panel is larger than that in the bottom 
panel by a factor of $\sim 1.5$. 
This model gives a poorer fit to the radio data than the geometrically thinner models.
In this case, although the black hole is rotating rapidly, both the empty and filled
models provide reasonably similar fits to the data. This is because the funnel emission
is somewhat suppressed relative to the other models and so the difference between
empty and filled funnels is less extreme.
}
    \label{fig:thickdisk7}
\end{figure}

To scale our GRMHD models to Sgr A*, we set the black hole mass to be 
$M=4\times 10^{6} M_\odot$ \citep{Gillessen+09} 
and adjust the mass accretion rate so that the resulting flux at $230$ GHz 
is roughly consistent with the observational data.
This emission likely originates from within a few Schwarzschild radii of the
supermassive black hole \citep{Doeleman+08}, a region which is well resolved by 
the GRMHD simulations and has reached a quasi-steady state.
In Figure~\ref{fig:thinnermad1} we show spectra from the thin-MAD 
model with a black hole spin of $a=0.1$, for 
two different observer inclinations of $\theta=\pi/2$ 
(perpendicular to the spin axis), and $\theta=\pi/3$.
The ``empty'' model corresponds to the case where the funnel material does not
contribute significantly to the observed spectra. 
In this case, we have removed all the plasma from the centre of the funnel and 
so the emission originates in the accretion disk and in the funnel wall.
The ``filled'' model corresponds to the extreme case where the funnel 
is filled with constant profiles of mass and internal energy densities.
The values are chosen to be equal to those at the edge of the funnel wall.

The radio data points and IR limits are the same as those considered by 
\citet{Chan+15a}. In particular, the IR limits represent the highest and lowest 
observed fluxes. The X-ray flux during quiescence is marked by the square data 
point \citep{Baganoff+03}. The diamond marks $10\%$ of the quiescent 
X-ray flux, which is the estimated contribution from the inner accretion flow 
\citep{Neilsen+13}. The range of observed X-ray flares is represented by the 
star and corresponding error bars \citet{Neilsen+13}.

The mass accretion rate in Figure~\ref{fig:thinnermad1}
is set such that the average rate at the horizon is 
$\dot{M}\approx 10^{-7} \dot{M}_\text{Edd}$.
Interestingly, the radio emission at frequencies $\nu\lesssim 10^{12}$ Hz
is not sensitive to the mass-loading of the funnel.
This is because this emission is dominated by the funnel wall. This is 
consistent with the findings of \citet{Moscibrodzka+14}, who refer to this 
region as the ``jet sheath''. 
Although there is a clear increase at 
IR and optical frequencies relative to the empty funnel case,
both the empty and filled funnel models are largely consistent with the data.

In Figure~\ref{fig:thinnermad5} we show the spectra for the higher-spin case of
$a=0.5$. The spectra are qualitatively similar to those in 
Figure~\ref{fig:thinnermad1}, however the enhancement at IR and optical
frequencies is larger. 
To obtain better fits with the filled model, we reduced the accretion rate
by a factor of $\sim 1.5$ in the bottom panel relative to the top panel.
The increase in this synchrotron component causes
a corresponding increase in synchrotron self-Compton emission in the hard 
X-rays. As with the $a=0.1$ model, both the empty and 
filled funnel cases fit the data reasonably well, however, the IR flux in the 
filled model is very close to the upper limits on the observed flux.

In Figure~\ref{fig:thinnermad9} we show the thin-MAD model with $a=0.9$. The
IR and X-ray limits clearly disfavour the case where the funnel material 
contributes significantly to the emission. 
Although the X-ray and IR emission can be brought within the limits by adjusting 
the mass accretion rate, this would also significantly reduce the radio flux
which originates in the funnel wall and which is independent of the 
mass-loading.
The difference between the empty and filled models is even more dramatic
in the extreme $a=0.99$ case, which we show in Figure~\ref{fig:thinnermad99hc}. 
As discussed in \citet{O'Riordan+16b}, the emission from this model is strongly 
dominated
by the near-horizon plasma. In order to give reasonable fits to the data, even
in the empty funnel case, we suppressed this near-horizon radiation by 
imposing a 
temperature ratio of $\Tcal=300$ on the inflowing material. A similar result was 
found by \citet{Chan+15a}, whose best-fit MAD models have very large 
proton-to-electron temperature ratios in the disk.

In Figure~\ref{fig:nonmad} we show the spectra calculated from our SANE model
with $a=0.92$ and a mass accretion rate of 
$\dot{M}\approx 10^{-6} \dot{M}_\text{Edd}$. 
This model has the same scale height of $H/R\approx0.2$ as our thin-MAD models.
As in the thin-MAD case, the radio emission is insensitive to the mass loading 
of the funnel.
The higher mass accretion rate results in a larger optical depth, which is 
clearly reflected in the high-energy parts of the spectra that show
multiple Compton scatterings.
Interestingly, as in the high-spin MAD models, the filled funnel 
model significantly over-produces IR emission and so an empty funnel is favoured 
by the data.

To conclude, for all our thin-MAD models and our SANE model, we find that the 
radio flux is dominated by the
funnel wall and is largely independent of the mass-loading of the jet.
We also find a significantly larger IR flux in the filled models than in the 
empty models. From this, we expect that the ratio of the IR flux and 230 GHz 
flux could be used as a probe of mass-loading processes in the funnel.
Furthermore, in the context of Sgr A*, although our low-spin models are consistent 
with the data in 
both empty and filled funnel cases, the higher-spin models only fit the data
provided the funnel material does not contribute significantly to the observed
spectrum.

In Figure~\ref{fig:thickdisk7} we show the spectra calculated from our thick-MAD 
model which has a black hole spin of $a=0.9375$ and a very geometrically-thick 
disk ($H/R\approx 1$). 
As in Figure~\ref{fig:thinnermad5}, the accretion rate in the bottom panel is 
$\sim 1.5$ times lower than that in the top panel.
Although the emission from our thin-MAD and SANE models is dominated by the 
region $r\lesssim 30 M$, which has reached a quasi-steady state, the 
outer radii of our thick-MAD model can contribute significantly to the emission.
The outer radii of our GRMHD models have not had enough time to reach a steady 
state and so the plasma properties depend strongly on the initial conditions in
the torus.
Furthermore, the 230 GHz flux which we have been using to normalise our models 
likely originates in the inner few $r_g$ of the accretion flow \citep{Doeleman+08}.
Therefore, we follow the procedure of \citet{Shcherbakov+12} to analytically 
extend the fluid quantities to large radii. 
We extend the fluid properties at $r=30 M$ as
power-laws out to the Bondi radius in order to match the estimated density
and temperature for Sgr A* at this radius. We further assume an isothermal jet
with electron temperature $\Theta=kT/mc^2=50$ 
\citep{Moscibrodzka+14,Chan+15a,Gold+17}, which provides a 
better fit to the radio emission than a constant temperature ratio for this 
model.
The difference between the empty and filled funnel models is smaller
than in the high-spin thin-MAD and SANE cases and so
both provide similar fits to the data.
Contrary to the previous cases, the funnel filling primarily affects the
lower-frequency emission. This is consistent with \citet{Gold+17}, who
found that the 230 GHz images of their models were affected by the funnel 
filling.
We will perform a more thorough investigation of the dependence on the disk 
scale height and prescriptions for extending the data to the Bondi radius in a 
future work.

\subsection{Predictions for Spectra of the Low/Hard State in XRBs}
\label{sec:xrb_spectra}

\begin{figure}
    \centering
    \includegraphics[width=\figfactortwo\linewidth]{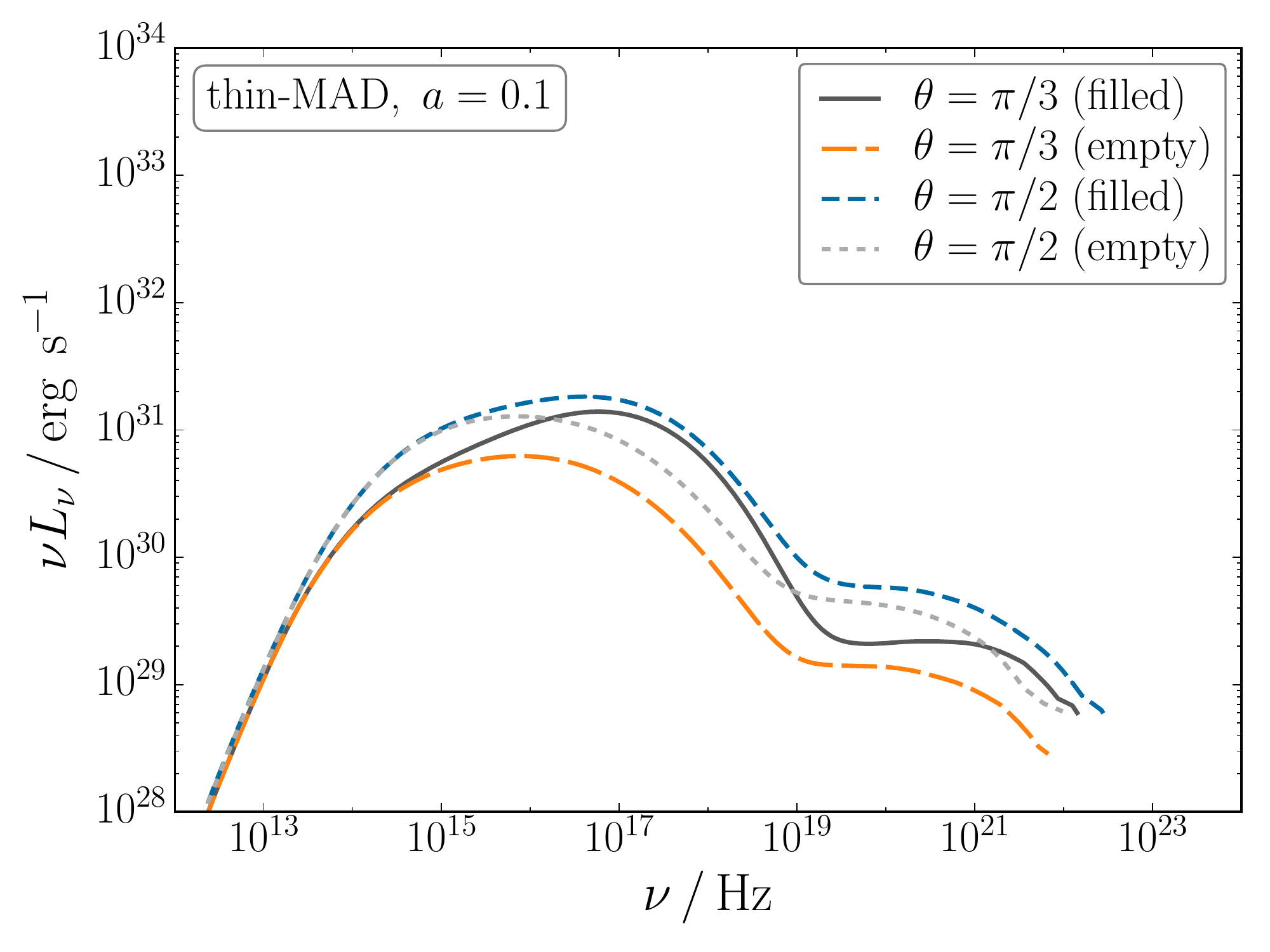}
    \includegraphics[width=\figfactortwo\linewidth]{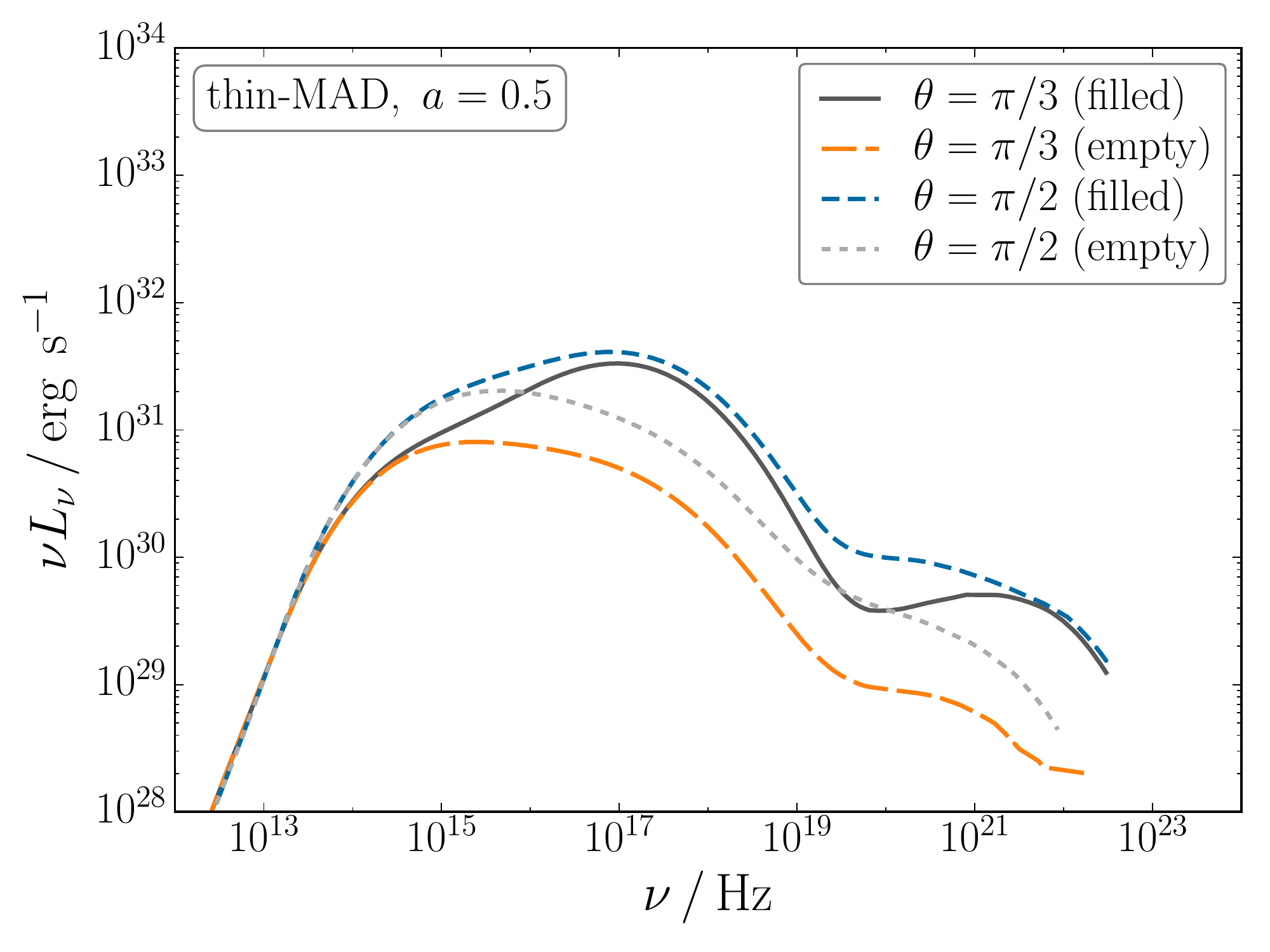}
    \caption{Spectra for thin-MAD models with $a=0.1$ (top), and $a=0.5$ (bottom).
    The black hole mass has been set to $M=10M_\odot$.
    The spectra are qualitatively similar to the results for Sgr A*.
    The optical and lower frequency emission is insensitive to the funnel material,
    while the hard UV and soft X-rays are significantly enhanced relative to the
    empty funnel case.
}
    \label{fig:thinnermad1_thinnermad5_XRB}
\end{figure}

\begin{figure}
    \centering
    \includegraphics[width=\figfactortwo\linewidth]{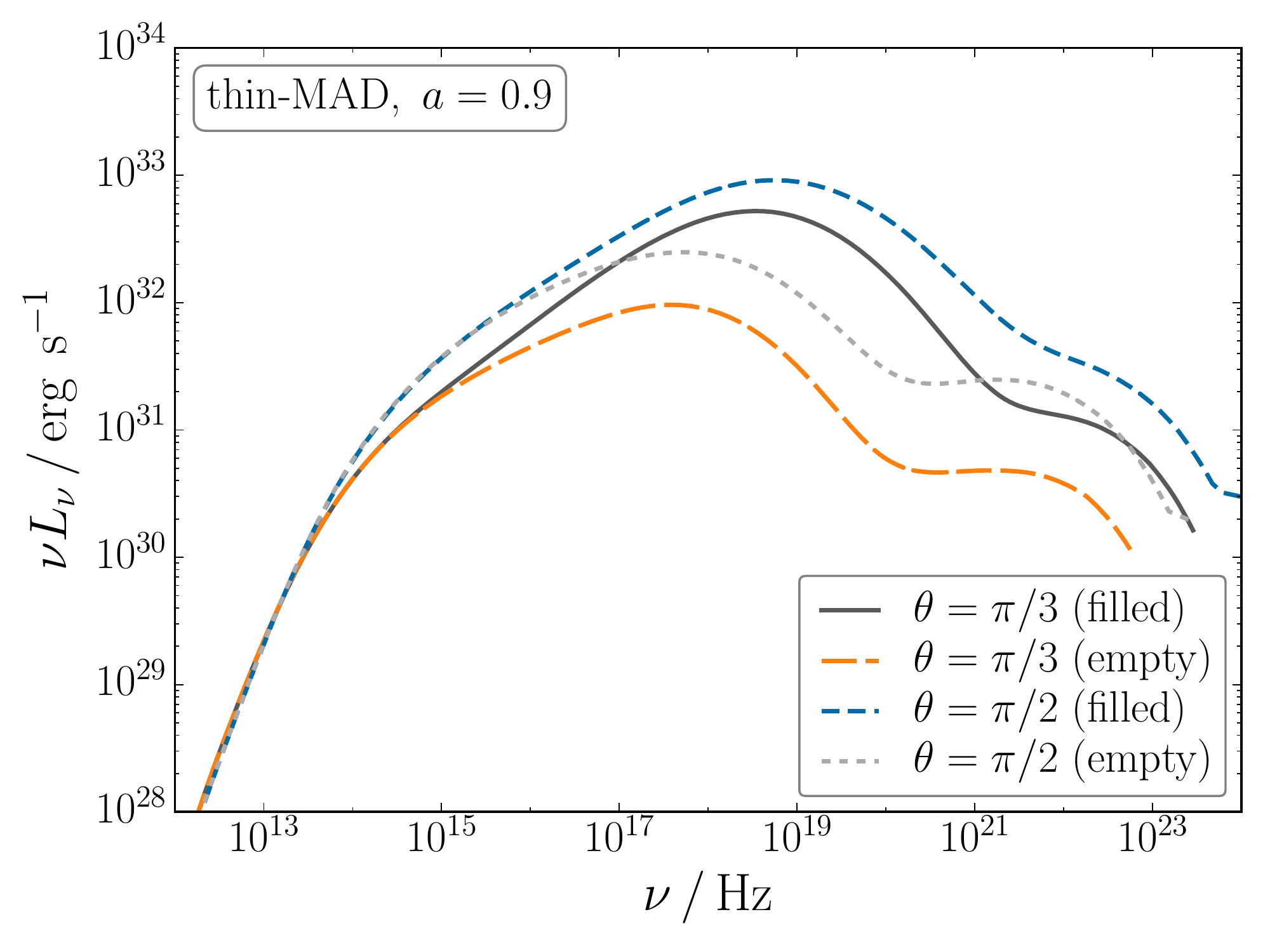}
    \includegraphics[width=\figfactortwo\linewidth]{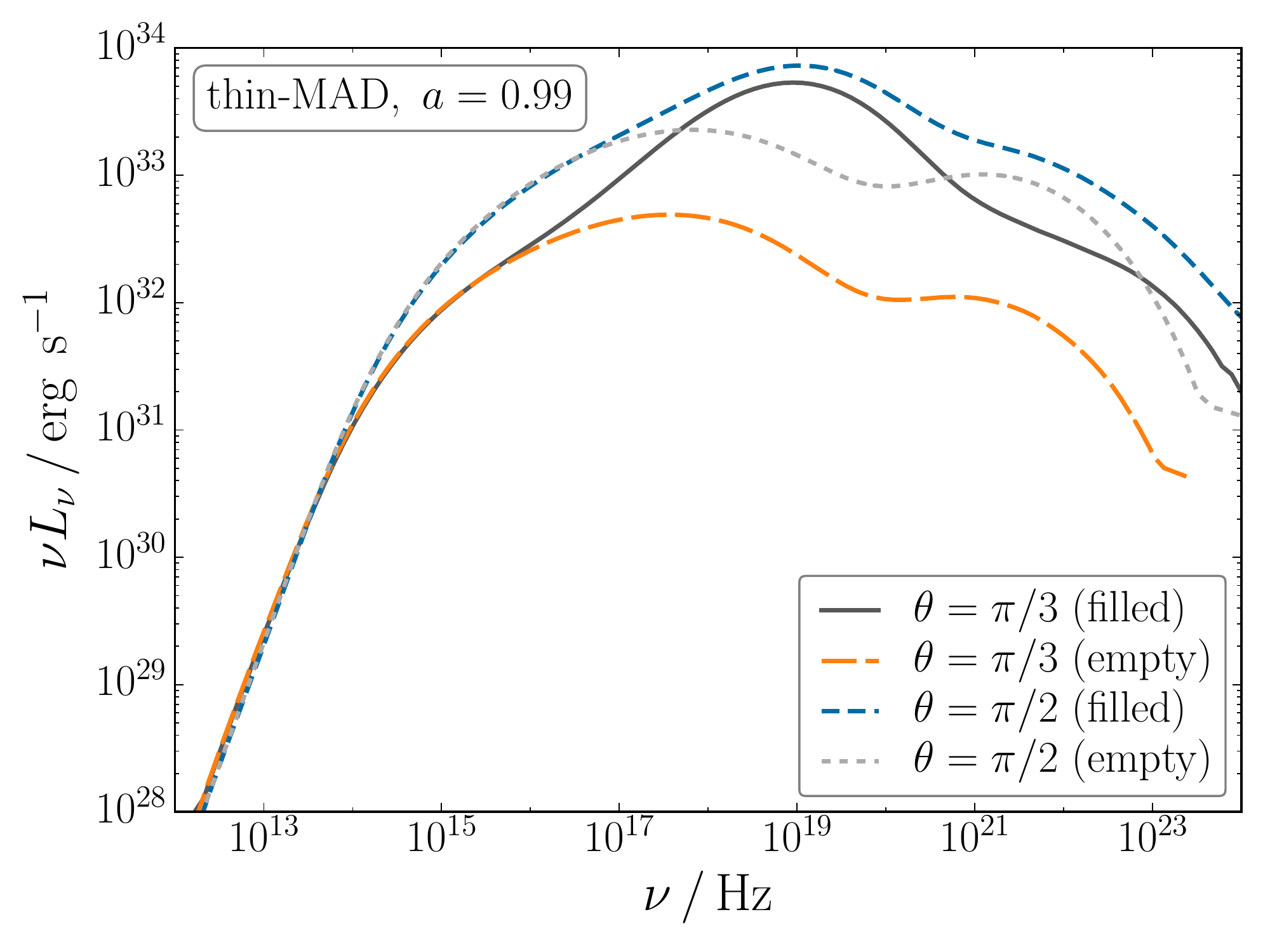}
    \caption{Spectra for the thin-MAD models with $a=0.9$ (top), and $a=0.99$ (bottom).
    The X-ray flux is significantly higher in the filled funnel models, while emission
    at frequencies $\lesssim 10^{16}$ Hz is unaffected by the funnel matter.
}
    \label{fig:thinnermad9_thinnermad99hc_XRB}
\end{figure}

\begin{figure}
    \centering
    \includegraphics[width=\figfactor\linewidth]{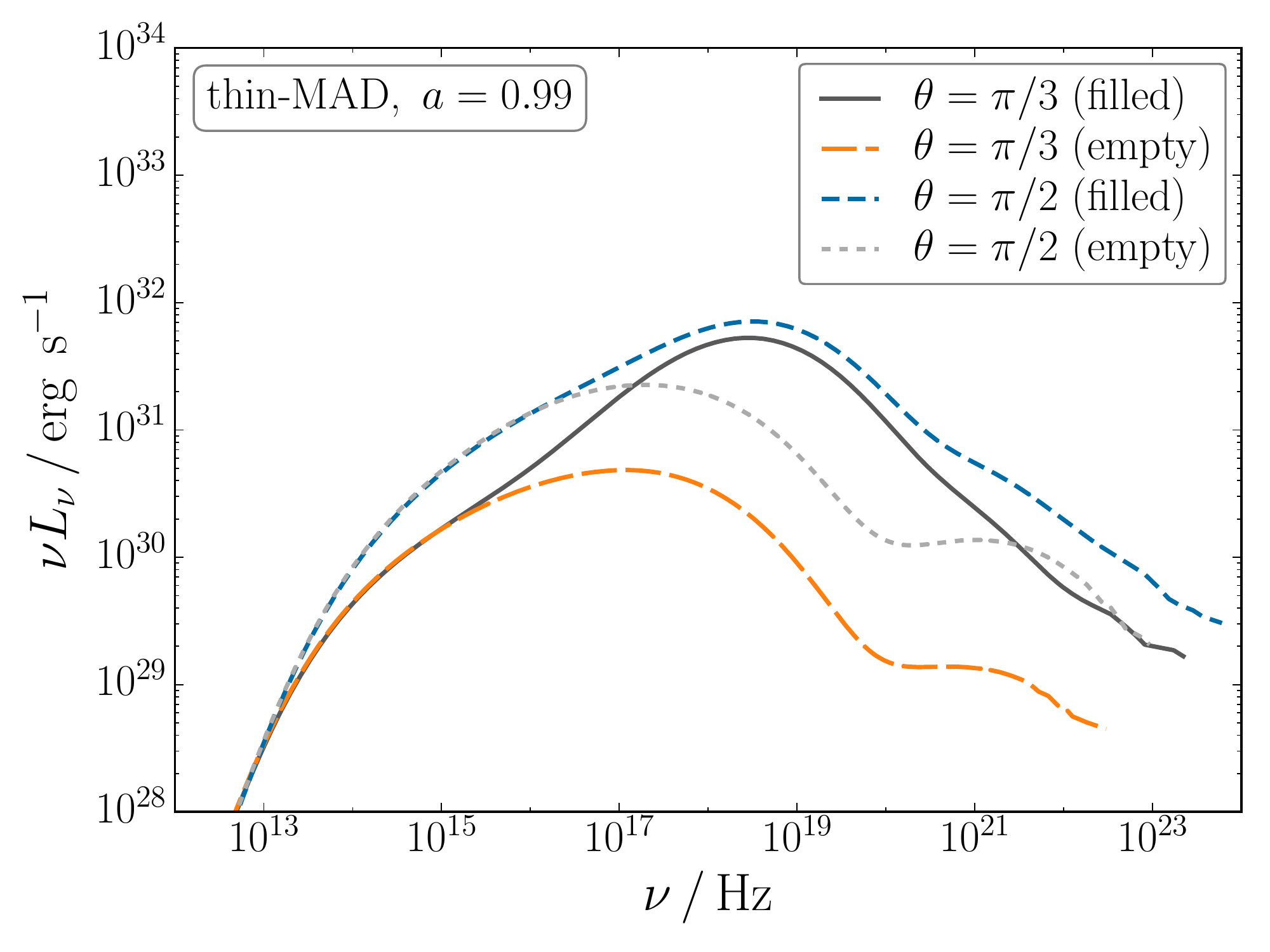}
    \caption{Same as the bottom panel of Figure~\ref{fig:thinnermad9_thinnermad99hc_XRB},
but with a lower accretion rate of $\dot{M}\approx 10^{-7}\dot{M}_\text{Edd}$.
Although the luminosity is significantly lower than in the previous case, the 
frequencies at which the emission is enhanced are similar. This is due to the 
reasonably weak dependence of the frequency on the accretion rate.
}
    \label{fig:thinnermad99hc_XRB_Mdot}
\end{figure}

In this section we scale our thin-MAD models to the low luminosity state in XRBs
by setting the black hole mass to $M=10M_\odot$. 
For comparing the different GRMHD models,
we fix the mass accretion rate to be
$\dot{M}\approx 10^{-6}\,\dot{M}_\text{Edd}$.
To maximize the potential effects of the funnel emission, we again consider the
case where the proton-to-electron temperature ratios in the disk and jet are 
$\Tcal_\text{disk}=30$ and $\Tcal_\text{jet}=3$.

In Figure~\ref{fig:thinnermad1_thinnermad5_XRB} we show the spectra for the 
low-spin models with $a=0.1$ and $a=0.5$. 
The results are qualitatively similar to the corresponding spectra for Sgr A*, 
with differences in the peak frequencies and overall luminosity due to changes 
in the black hole mass and accretion rate.
In particular, we find that the filled funnel models show enhanced 
hard UV/soft X-ray emission, while the optical and lower-frequency fluxes are 
unaffected by the mass-loading.

In Figure~\ref{fig:thinnermad9_thinnermad99hc_XRB} we show the spectra for the
high-spin models with $a=0.9$ and $a=0.99$.
We find very large differences between the empty and filled funnel models, with 
the funnel contribution shifting to higher frequencies.
In this case, the X-rays and $\gamma$-rays are significantly modified by the 
funnel matter,
while frequencies below $\sim 10^{16}$ Hz are unaffected by the funnel 
contribution.
In the $a=0.99$ case, the radiative efficiency is large, approaching values
$\gtrsim 10\%$, especially in the filled funnel model.
A similar result was reported by \citet{Ryan+17}, who found that accretion flows
with $a=0.5$ can approach $1\%$ radiative efficiency by 
$\dot{M}\sim 10^{-5} \dot{M}_\text{Edd}$.
To avoid complications due to radiative cooling, we investigate a lower 
accretion rate of 
$\dot{M}\approx 10^{-7}\dot{M}_\text{Edd}$,
and show the resulting spectra in Figure~\ref{fig:thinnermad99hc_XRB_Mdot}.
The spectra in the hard X-rays and below 
are qualitatively similar to those in 
Figure~\ref{fig:thinnermad9_thinnermad99hc_XRB}, and so our conclusions about 
the effects of the funnel mass-loading still hold.
This is not surprising since, 
as shown in Appendix~\ref{sec:M_and_Mdot}, although the luminosity depends very 
strongly 
on the accretion rate $L_\text{syn}\sim \dot{M}^2$, the frequency depends 
only weakly on $\dot{M}$ as $\nu_\text{syn}\sim \dot{M}^{1/2}$.
There is a larger difference in the synchrotron self-Compton component due to 
the linear dependence of the Compton $y$ parameter on $\dot{M}$ 
(see Appendix~\ref{sec:M_and_Mdot}).

\section{Summary and Discussion}
\label{sec:discussion}
In this work, we investigated the observational effects of mass-loading
in BZ jets.
We considered the case in which the plasma in the funnel remains 
force-free, which allowed us to treat the mass-loading in a post-processing 
step.
We found significant differences between models with an empty funnel and
models where the funnel was filled with highly-magnetized plasma.
In particular, in the context of Sgr A* the IR and optical flux is enhanced 
relative to the empty funnel case.
Interestingly, the radio emission 
from our thin-MAD and SANE models is 
dominated by the funnel wall and so is largely unaffected by the mass-loading.
We identify the ratio of the IR and 230 GHz flux as a 
potential observational 
probe of the filling factor of the funnel. 

As argued by \citet{Gold+17}, understanding the contribution from the funnel
material will be extremely important for interpreting future EHT observations 
of the black hole shadow in Sgr A*. They showed that the absence of
significant 230 GHz emission from the funnel can appear as a ``hole'' in
the images, mimicking features of the black hole shadow.
Since the radio emission from our models is not affected by the funnel material, 
we expect that the mass-loading of the BZ jet will not have a large
impact on images from the EHT (unless the disk is very thick with $H/R\sim1$,
as shown in Figure~\ref{fig:thickdisk7}). This means that even mass-loaded BZ 
jets may appear as ``holes'' in images from the EHT.

We find qualitatively similar results 
in the context of XRBs, although shifted to higher frequencies due to changes in 
the black hole mass and accretion rate.
It is often argued that inverse Compton emission from 
a corona of hot electrons surrounding the inner accretion flow
is responsible for the X-ray emission observed in XRBs
\citep[e.g.,][]{Titarchuk94,MZ95,Gierlinski+97,Esin+97,Esin+01,Poutanen98,
    CB+06,Yuan+07,NM08,Niedzwiecki+14,Niedzwiecki+15,QL15}.
However, there is significant degeneracy between these models and 
ones in which a large fraction of the X-ray
emission originates in the base of the jet
\citep[e.g.,][]{MiRo94,Markoff+01,Markoff+03,Markoff+05,FKM04,BRP06,Kaiser06,
    GBD06,Kylafis+08,Maitra+09,PC09,PM12,Markoff+15,O'RIordan+16a}.
Understanding the funnel mass-loading could be crucial for breaking this 
degeneracy and constraining
the role of the jet in producing the observed high-energy X-ray emission
in the low/hard state.

Our results have interesting implications for explaining the
scatter in the fundamental plane of black hole activity
\citep{Merloni+03,FKM04}.
The fundamental plane is an empirical correlation between black hole mass, radio
luminosity, and X-ray luminosity which spans the mass scale from XRBs to active
galaxies. This correlation suggests that low-luminosity accreting black hole 
systems are scale invariant.
Our results imply that differences in the jet mass-loading could contribute to 
the scatter about the best-fit correlation.
In particular, at high black hole spin the X-ray emission can vary by more than 
two orders of magnitude between the empty and filled models, while the radio
emission remains constant.
Therefore, in addition to variations in quantities such as the mass accretion 
rate, black hole spin, and viewing angle, the mass-loading of the jet could 
play a significant role in producing the observed scatter.

For our empty funnel models, we set the plasma density in the funnel to 
zero. However, this case represents a wider class of models in which the funnel 
contains material that does not contribute significantly to the emission.
For example, models in which
the proton-to-electron temperature ratio in the jet is comparable to that in the
disk result in similar spectra to the empty funnel cases. This is because the 
denser funnel wall dominates the jet component unless the plasma in the centre 
of the funnel is hot enough.
For similar assumptions about the electron temperatures, the spectra from more 
complicated matter profiles in force-free jets should fall within the
limits considered here. 
In a future work, we will investigate observational signatures of the regime 
where the force-free approximation breaks down. 
As shown by \citet{GL13}, in this case the solution in the funnel can deviate 
significantly from the BZ funnel solutions in our dynamical models.

We have not specified a mass-loading mechanism, but have simply 
compared spectra from the empty funnel case to the extreme case of a steady, 
force-free funnel with constant mass and internal energy density profiles.
As well as spectral properties, we expect that variability studies will play a 
key role in constraining the mass-loading physics systems such as Sgr A*.
Importantly, many of the proposed mass-loading mechanisms operate on very 
different time-scales, and so could in principle be distinguished by the EHT.
For example, pair production by vacuum gaps in the black hole magnetosphere
is expected to be intermittent, and vary on timescales comparable to the 
light-crossing time of the black hole \citep{LR11,BT15}.
This timescale is extremely short in Sgr A*,
roughly equal to one minute. 
However, it might be possible with the EHT to study structures in the accretion 
flow that vary on minute timescales \citep{Doeleman+09,Doeleman+2009}.
This could provide valuable constraints on the physics of near-horizon 
mass-loading.
Other mass-loading processes may operate on timescales
significantly longer than the light-crossing time.
For example, pairs may be produced by photon annihilation 
\citep{Moscibrodzka+11} on timescales determined by radiation field of the disk.
Furthermore, magnetohydrodynamic processes such as magnetic Rayleigh-Taylor 
instabilities in the funnel wall \citep[][and Appendix~\ref{sec:filling}]{MTB12}, 
or magnetic field polarity 
inversions in the disk \citet{Dexter+14} can inject matter from the disk into 
the centre of the funnel. These processes operate on spatial scales much larger
than the Schwarzschild radius, and so the corresponding variability could be 
resolved by future observations.

A significant limitation of the current work is our simplified treatment of the
emitting electrons. In particular, we neglect the contribution from non-thermal
electrons which might be present due to acceleration by shock waves 
\citep[e.g.,][]{Sironi+15}, magnetic reconnection \citep[e.g.,][]{SS14},
or due to the injection process itself \citep[e.g.,][]{LR11}. 
Although these non-thermal electrons would likely affect the high-frequency 
emission in our spectra, including these processes would introduce additional
poorly-constrained free parameters into our models, and so
we neglect this contribution as a first step.
We also use a simple prescription for calculating the electron temperature by 
varying the proton-to-electron temperature ratio as a function of the 
plasma $\beta$. 
This ratio is a free parameter which is poorly constrained both by theory and 
observations.
We choose values consistent with the findings of recent, sophisticated 
models of the electron thermodynamics in collisionless accretion flows
\citep{Ressler+15,Foucart+16,Sadowski+17a}, which show that 
the electron temperature is comparable to the proton temperature in 
highly-magnetized regions of the flow.
Modelling the electron physics in accretion disks and jets remains an active 
area of research, which will hopefully be informed further by upcoming 
observations with the EHT.

\acknowledgments 
The authors would like to thank Alexander Tchekhovskoy for providing simulation data.
AP and MOR acknowledge the DJEI/DES/SFI/HEA Irish Centre for High-End Computing 
(ICHEC) for the provision of computational facilities under project ucast008b.
MOR is supported by the Irish Research Council under grant number GOIPG/2013/315.
This research was partially supported by the European Union Seventh Framework 
Programme (FP7/2007-2013) under grant agreement no 618499.
JCM acknowledges NASA/NSF/TCAN (NNX14AB46G), NSF/XSEDE/TACC (TGPHY120005), and 
NASA/Pleiades (SMD-14-5451).
The authors would like the thank the anonymous referee for constructive comments
which have improved the quality of the manuscript.
\software{\texttt{HARM} \citep{Gammie+03}, \texttt{grmonty} \citep{Dolence+09}}

\appendix
\numberwithin{equation}{section}
\setcounter{figure}{0}
\section{Mass and Internal Energy Densities in MAD and SANE GRMHD Models}
\label{sec:filling}
\begin{figure}
    \centering
\minipage{0.5\textwidth}
    \includegraphics[width=\figfactorfour\linewidth]{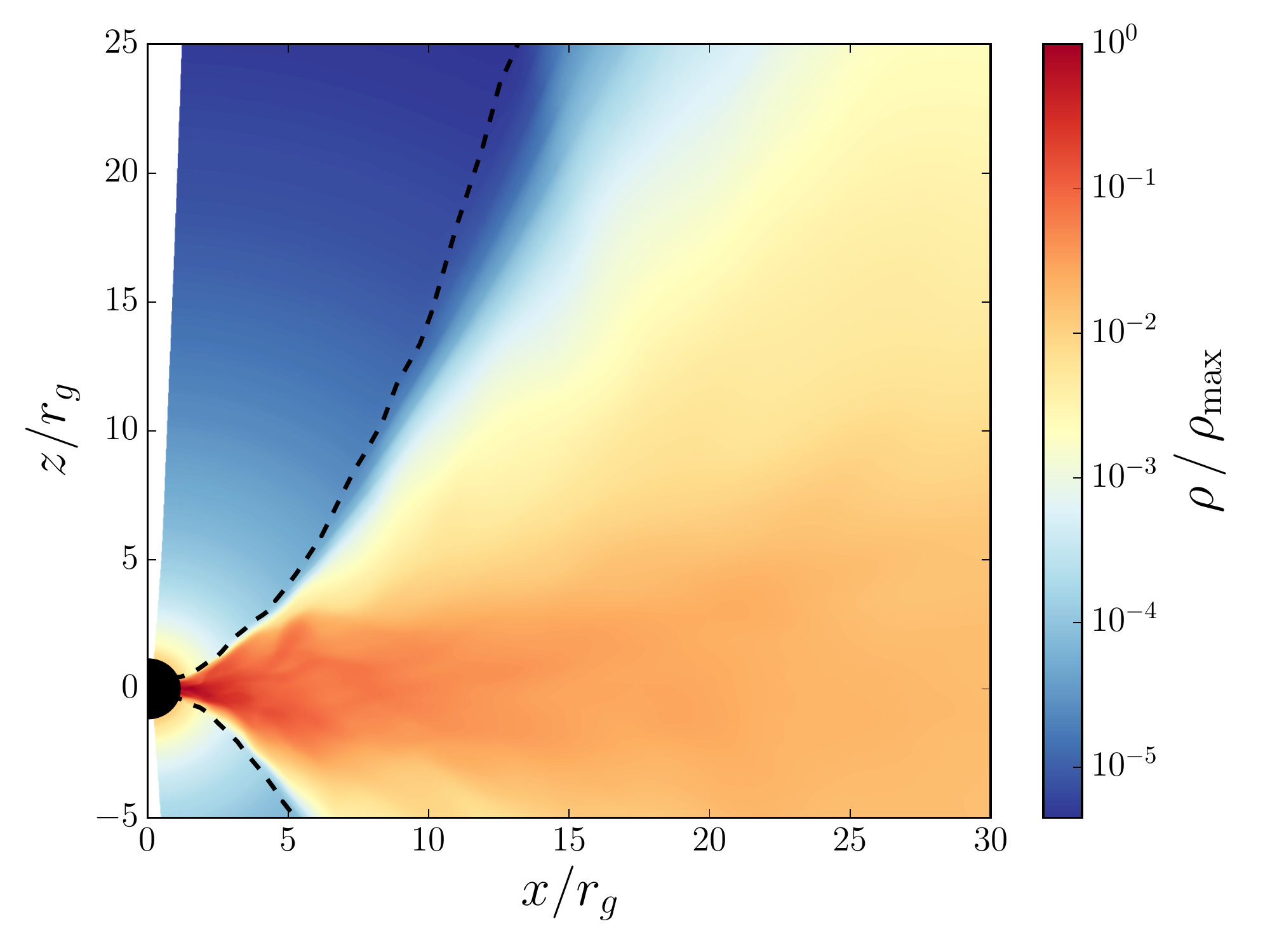}
    \includegraphics[width=\figfactorfour\linewidth]{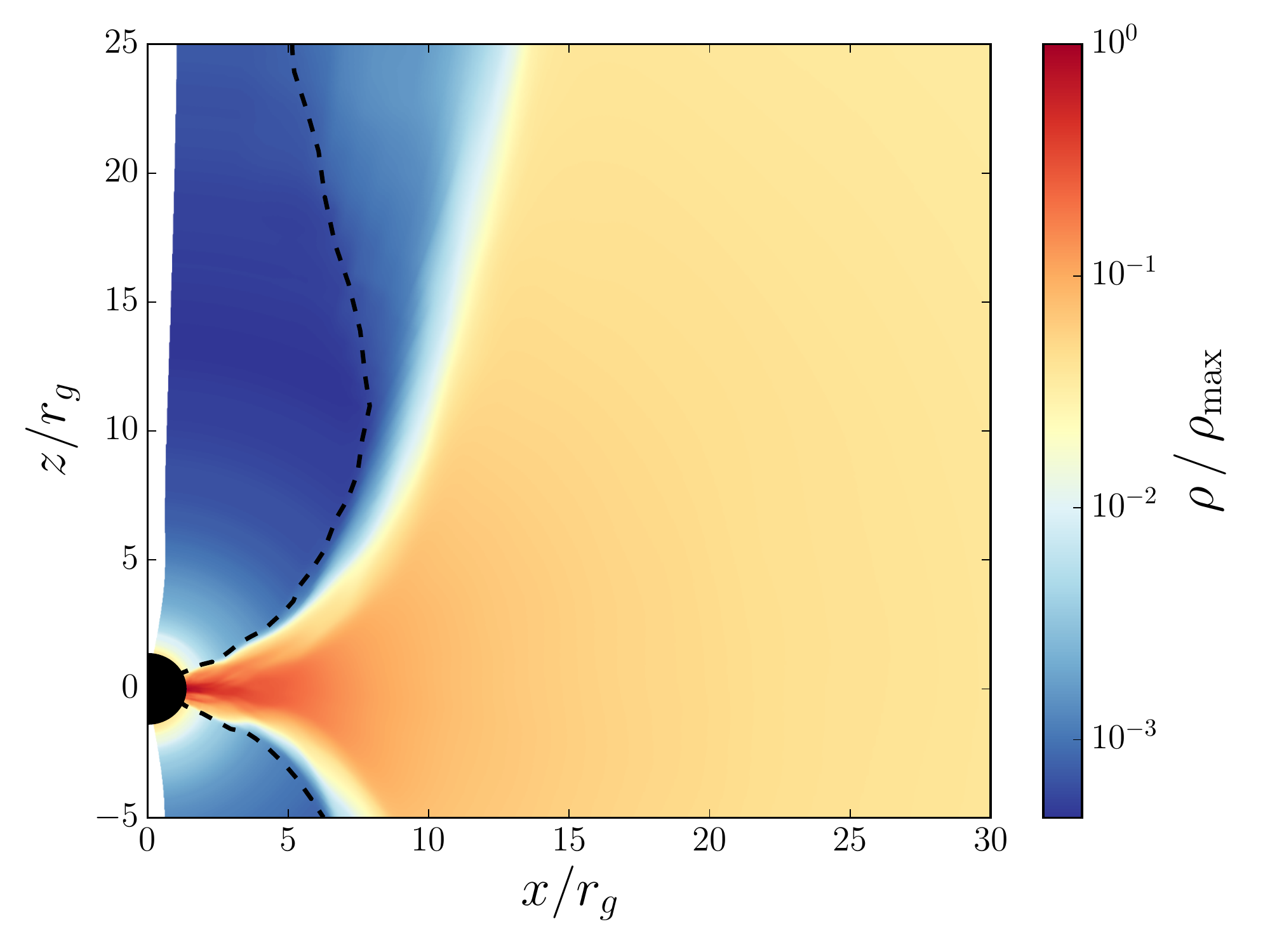}
    \includegraphics[width=\figfactorfour\linewidth]{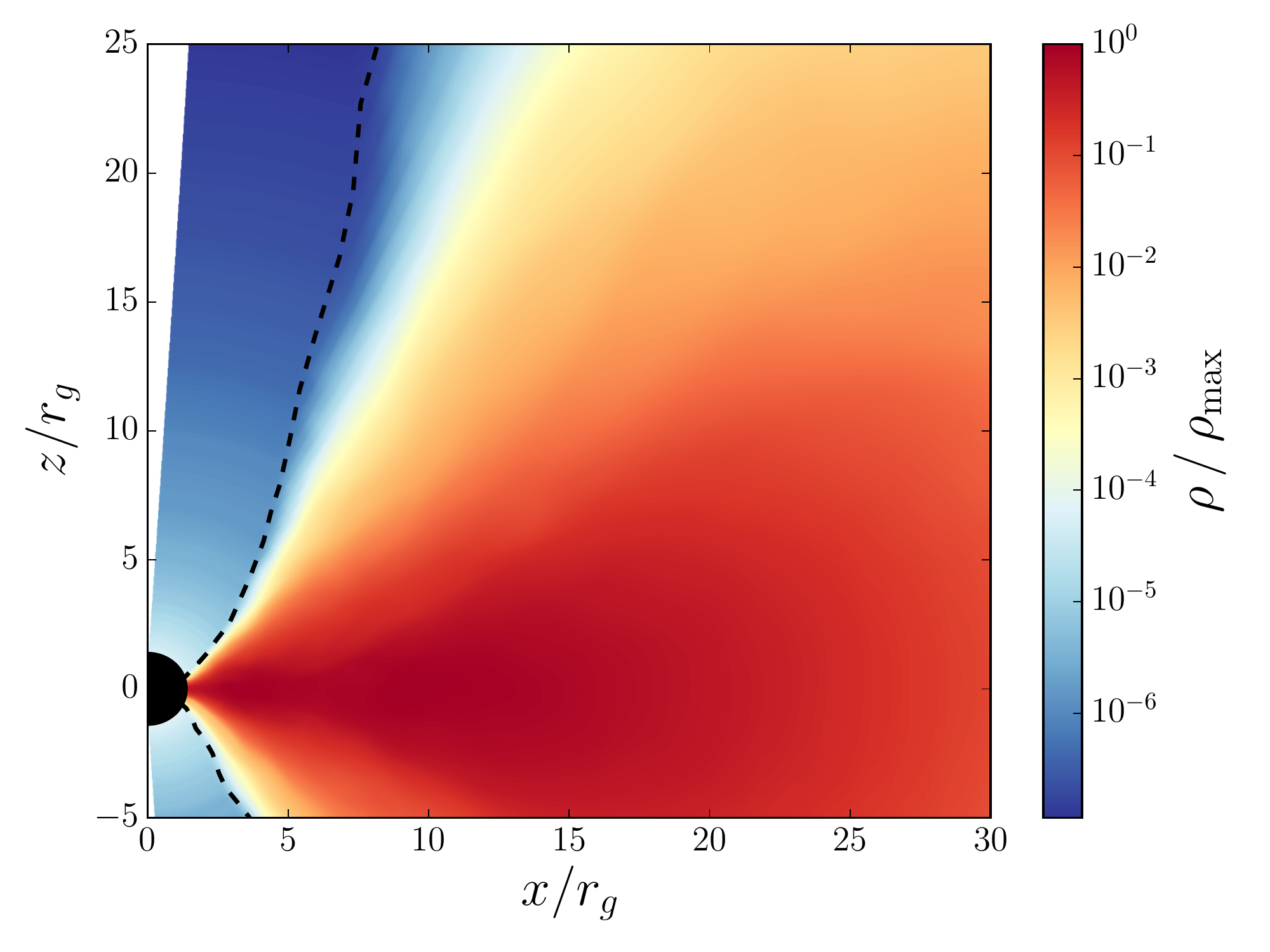}
\endminipage\hfill
\minipage{0.5\textwidth}
    \includegraphics[width=\figfactorfour\linewidth]{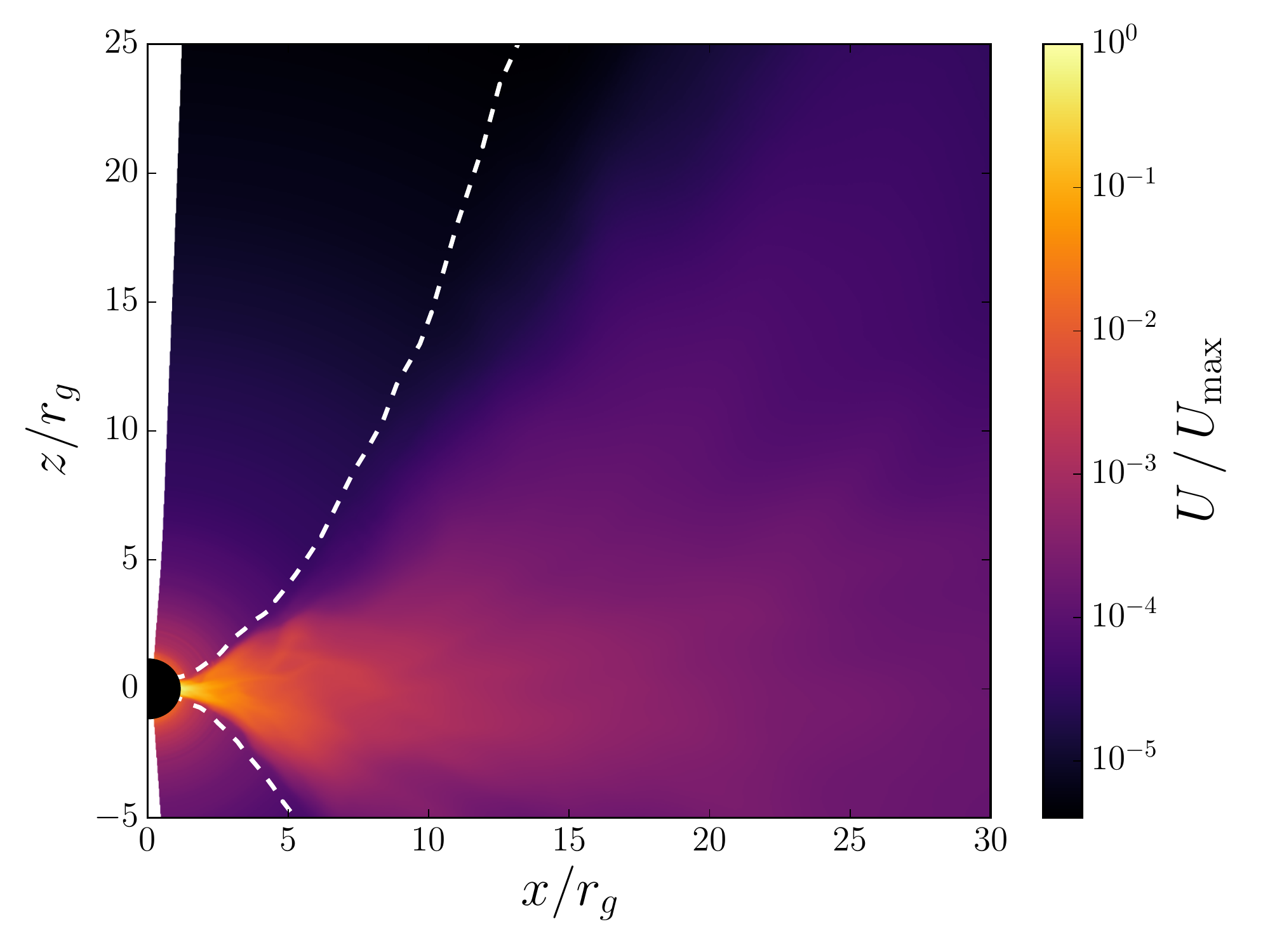}
    \includegraphics[width=\figfactorfour\linewidth]{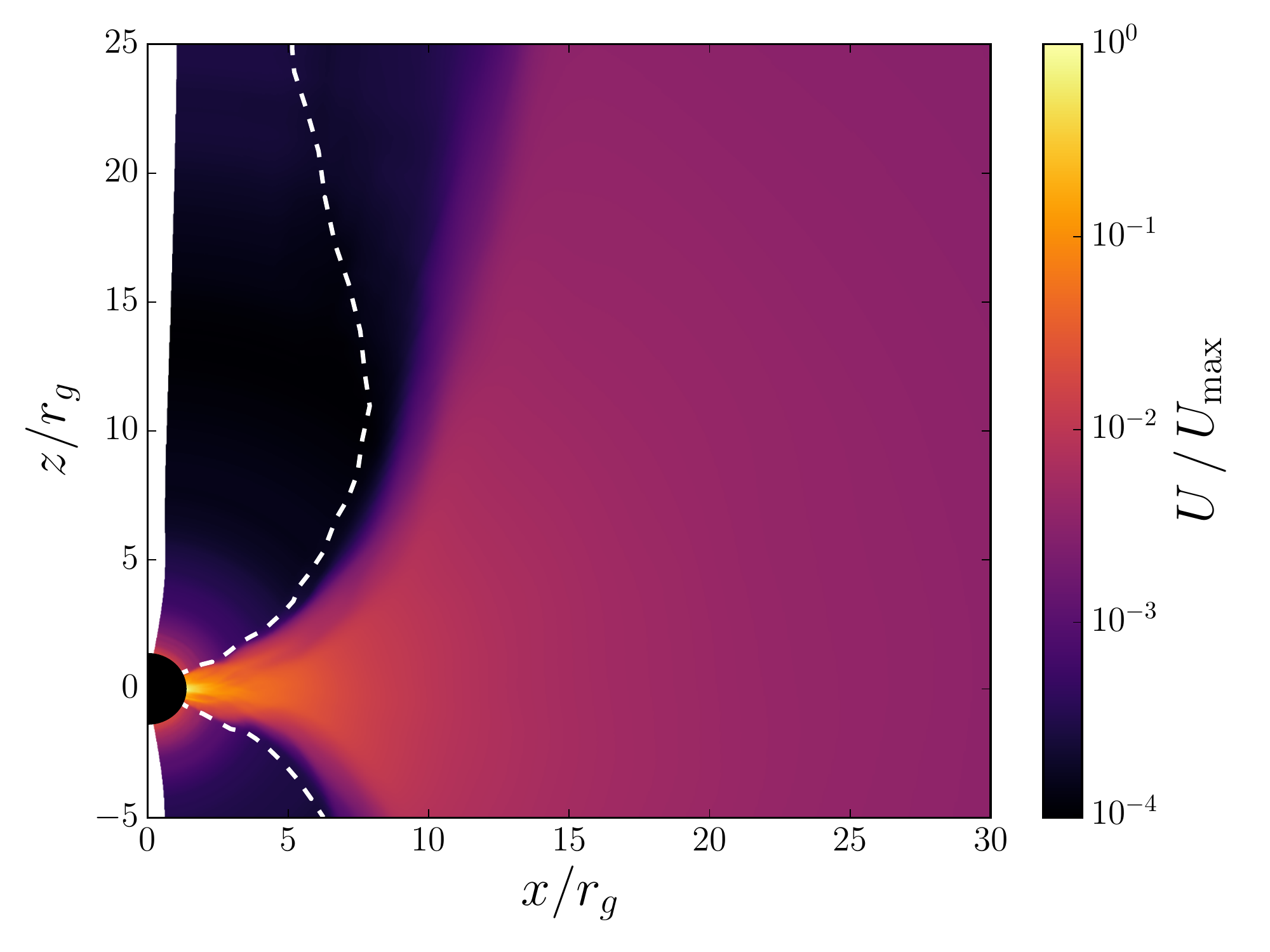}
    \includegraphics[width=\figfactorfour\linewidth]{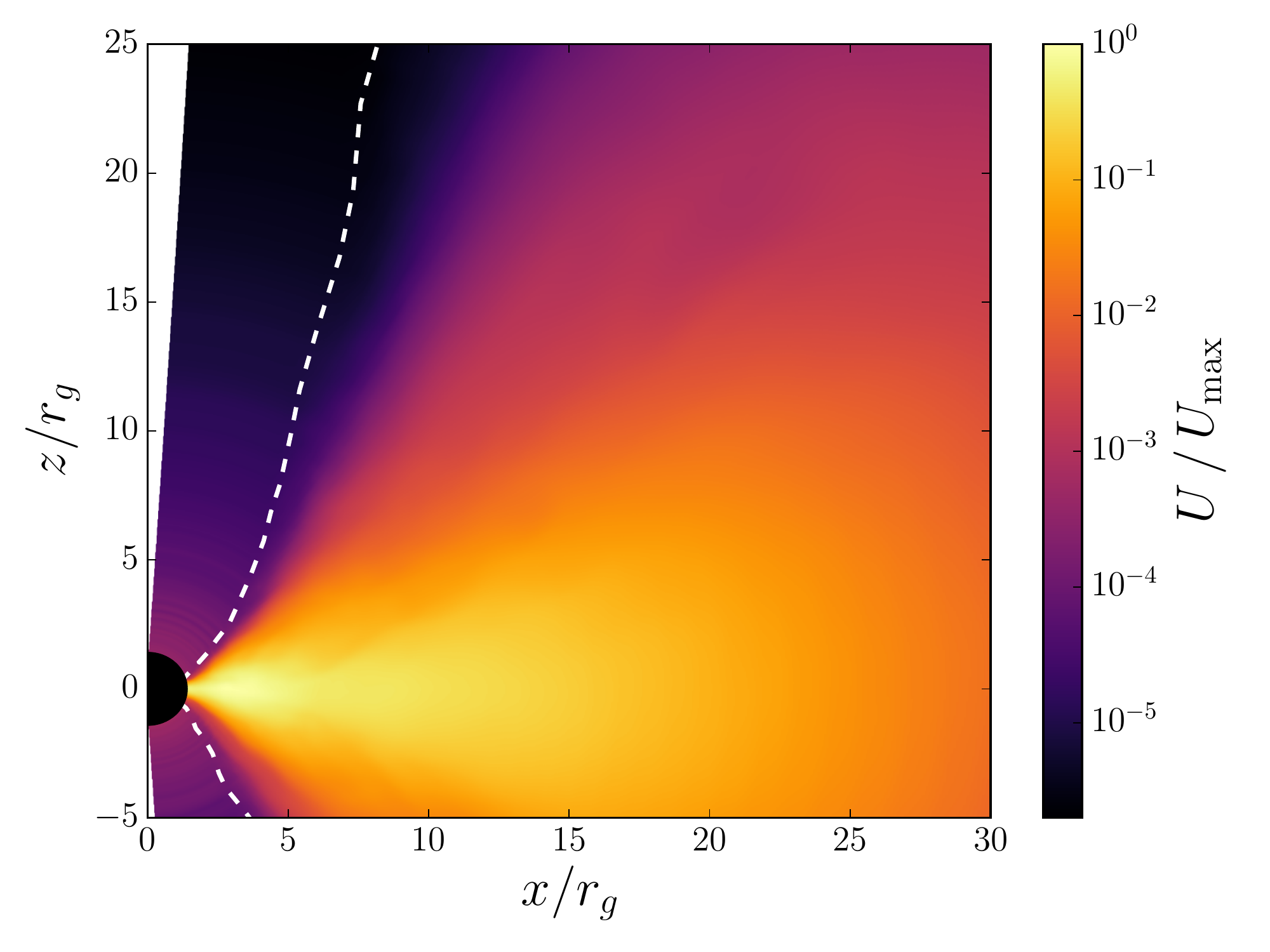}
\endminipage\hfill
    \caption{Snapshots of our MAD and SANE GRMHD models. 
       The left panels show the mass density, and the right panels show the
       internal energy density.
       The top panel shows the thin-MAD model with $H/R\approx 0.2$ 
       and $a=0.99$. The
       middle panel shows the thick-MAD model with $H/R\approx 1$ and $a=0.9375$. 
       The bottom panel shows the SANE model with $H/R\approx 0.2$ and $a=0.92$.
       The funnel regions are manually
       filled with constant profiles of mass and internal energy densities,
       according to the prescription described in Section~\ref{sec:models}. 
       The dashed lines represent the region which is removed in the ``empty''
       funnel models. In the text, we refer to the surface represented by the 
       dashed lines as the ``edge'' of the funnel wall.
}
    \label{fig:harm_models_filled}
\end{figure}
In Figure~\ref{fig:harm_models_filled} we show ($\phi$-averaged) snapshots of 
our MAD and SANE GRMHD models. 
The left panels show the mass density, and the right panels show the
internal energy density.
As in Figure~\ref{fig:harm_models},
the top panel shows the thin-MAD model with $H/R\approx 0.2$ 
and $a=0.99$, the
middle panel shows the thick-MAD model with $H/R\approx 1$ and $a=0.9375$,
and the bottom panel shows the SANE model with $H/R\approx 0.2$ and $a=0.92$.
In these plots, the funnel regions have been
filled with constant profiles of mass and internal energy
according to the prescription described in Section~\ref{sec:models}. 
The dashed lines represent the regions affected by the numerical floor material
(prior to the manual filling of the funnel),
which are removed in the ``empty'' funnel models. 
The jet in the thick-MAD model (middle panel) has a region near
$r\approx 20\, r_g$ which is not affected by the numerical density floors.
Instead, this is material which has moved from the disk into the funnel due to 
instabilities at the jet-disk interface \citep{MTB12}.
This is a transient feature, which has little effect on the 
spectra in this case. However, such disk-jet instabilities are a possible 
physical mechanism for mass-loading the jet.

\section{Dependence of the Spectra on the Black Hole Mass and Mass Accretion Rate}
\label{sec:M_and_Mdot}
The synchrotron luminosity scales with the fluid properties as
$L_\text{syn}\sim\rho B^2 \Theta^2 V$, where $\rho$ is the mass density, $B$ is 
the magnetic field strength, $\Theta=kT/mc^2$ is the electron temperature, and 
$V$ is the volume of the emitting region.
The mass density scales with the black hole mass and accretion rate as 
$\rho\sim \dot{M} t_g / V \sim \dot{M} / M^2$, 
where we have used that $t_g=r_g/c\sim M$ and $V\sim M^3$.
The magnetic energy density scales in the same way.
Since we are neglecting radiation pressure, the electron temperature
is simply proportional to the ratio of the internal and mass energy densities 
and so is independent of $M$ and $\dot{M}$.
Therefore, the luminosity scales as $L_\text{syn}\sim \dot{M}^2 / M$.
It is convenient to write the accretion rate as a fraction $\eta$ of the 
Eddington rate $\dot{M}_\text{Edd}$. Since $\dot{M}_\text{Edd}$ is 
proportional to the black hole mass, we find that
$\rho\sim\eta/M$ and so $L_\text{syn}\sim \eta^2 M$.
We can follow the same procedure to find scalings for the
synchrotron frequency
$\nu_\text{syn} \sim B \Theta^2 \sim \sqrt{\eta/M}$,
the optical depth $\tau = n \sigma_T R \sim \eta$, and 
the Compton $y$ parameter $y=16\Theta^2\tau \sim \eta$ \citep{RL79}.
We conclude that the luminosities of the synchrotron and Compton spectral 
components depend
strongly on the mass accretion rate as 
$L_\text{syn}\sim \dot{M}^2$ and 
$L_\text{Compton} = y L_\text{syn} \sim \dot{M}^3$, while the 
frequencies of these components depend only weakly on $\dot{M}$
as $\nu_\text{syn}\sim \dot{M}^{1/2}$ and 
$\nu_\text{Compton} \sim \Theta^2 \nu_\text{syn}\sim \dot{M}^{1/2}$.
Although we have neglected synchrotron self-absorption in these simple 
analytic scalings, we include this process in our numerical calculations 
of the spectra.



\bibliographystyle{apj}
\bibliography{jet_fill}

\end{document}